# Altered Cosmic Organics as Venus' Ultraviolet Absorbers

Rakesh Mogul[a,b,*], Mikhail Yu. Zolotov[c], Michael J. Way[d,e], Sanjay S. Limaye[f]

**Affiliations:**
[a]Chemistry & Biochemistry Department, California State Polytechnic University, Pomona; Pomona, CA, 91767, USA
[b]Blue Marble Space; Seattle, WA, 98104, USA
[c]School of Earth and Space Exploration, Arizona State University; Tempe, AZ, 85287, USA
[d]NASA Goddard Institute for Space Studies; New York, NY, 10025, USA
[e]Theoretical Astronomy Department of Physics and Astronomy, Uppsala University; Uppsala, SE-75120, Sweden
[f]University of Wisconsin, Madison; Madison, WI, 53706, USA

*Correspondence to: Rakesh Mogul (rmogul@cpp.edu)

**Keywords:**
- Venus, UV absorbers, cosmic dust, polycyclic aromatic hydrocarbons, iron

**Highlights**
- Venus' UV absorbers may be altered cosmic organic and iron-bearing compounds.
- Polycyclic aromatic hydrocarbons could be the primary absorbers across the cloud tops, aerosols, and sub-cloud atmosphere.
- Cosmically derived iron-bearing compounds could be dominant absorbers by mass.
- Geologically short timelines of cosmic influx would yield the necessary carbon and iron abundances.

## Abstract

Venus' ultraviolet (UV) absorbers significantly contribute to the atmosphere's energy budget. However, the composition of these absorbers remains a mystery. Here, we show that mixtures of polycyclic aromatic hydrocarbons (PAHs) and iron-bearing compounds, analogs of altered cometary dust, excellently match Venus' spectra from the cloud tops to the sub-cloud atmosphere across the UV and visible wavelengths. The molecular compositions for the cloud tops (5–10 ring PAHs and ferric chloride), decomposed cloud aerosols (3–4 ring PAHs and acid ferric sulfate), and sub-cloud atmosphere (5–10 PAHs and ferric hydroxy sulfates) are consistent with the stepwise alteration of cosmic dust. These steps include sourcing of the PAHs and iron from thermally unablated and ablated dust particles, respectively, reactions with sulfuric acid in the clouds, and thermal decomposition below the clouds. Geologically short timelines of cosmic influx (≥600 and ≥3 kyr) would respectively yield the cloud top carbon and iron concentrations. Hence, we propose a unified origin for Venus' absorbers, which may arise from cometary dust via altitude-dependent alteration pathways from the mesosphere to the surface, with PAHs serving as the primary UV absorbers and iron compounds as the dominant absorbers by mass.

## 1. Introduction

The brightness of Venus in the night sky has inspired observation for millennia. This brightness arises from the clouds, which contain chemicals and particles that absorb, reflect, and emit light. At the cloud tops are enigmatic molecules that absorb a majority of the deposited solar ultraviolet (UV) radiation (Esposito, 1980; Na et al., 1990; Titov et al., 2013; Pérez-Hoyos et al., 2018), thereby contributing directly to the planet's radiative energy budget and influencing atmospheric superrotation (Crisp and Titov, 1997; Limaye et al., 2018a). Venus' cloud aerosols (Bertaux et al., 1996) and sub-cloud atmosphere (Moroz et al., 1979; Ignatiev et al., 1997) also contain UV and blue absorbers. The identities of these total absorbers, however, remain unknown, despite decades of observations. Here, we present evidence that organic compounds of cosmic origin could be the primary absorbers across the cloud tops, aerosols, and sub-cloud atmosphere.

For Venus' cloud tops (nominally ~70 km), measurements of the UV/blue spectra (200–500 nm) include those from MESSENGER (Pérez-Hoyos et al., 2018), Venus Express (Lee et al., 2015), Pioneer Venus Orbiter (Esposito, 1980), and International Ultraviolet Explorer (Na et al., 1990). The absorbance in the far-UV (~200–230 nm) has been attributed to $SO_2$ (Esposito, 1980; Na et al., 1990), while differing absorbers in the near-UV (~300–400 nm) have been proposed. The presence of multiple absorbers has also been suggested (Pérez-Hoyos et al., 2018), since individual candidates incompletely match the full absorbance profile (~300–400 nm). Candidates based on spectral matches include OSSO, cyclic $S_2O$, iron compounds, and others. However, chemical modeling (Egan et al., 2025a) and experimental studies (Frandsen et al., 2020) show that OSSO and cyclic $S_2O$ are respectively too low in abundance or unstable to account for the observed absorption, while open-chain $S_2O$ yields poor spectral matches (Frandsen et al., 2020).

Several studies (Zasova et al., 1981; Petrova, 2018; Jiang et al., 2024; Egan et al., 2025b) point to the iron-bearing compounds as the cloud top UV absorbers. Significant near-UV matches are observed with solutions of ferric chloride in hydrochloric acid (HCl) and mixtures of ferric chloride and ferric sulfate in sulfuric acid ($H_2SO_4$) and HCl (Egan et al., 2025b). Near-UV matches are also observed with mixtures of solid ferric sulfates (rhomboclase and acid ferric sulfate) (Jiang et al., 2024). Solutions of ferric chloride in concentrated $H_2SO_4$, which likely contained ferric chloride and sulfates, reasonably match Venus' albedo (Zasova et al., 1981) and glory (Petrova, 2018). Organic molecules have also been suggested as the cloud top UV absorbers (Bertaux et al., 1996; Limaye et al., 2018b; Pérez-Hoyos et al., 2018; Spacek et al., 2024). Croconic acid was considered but shown to yield poor spectral matches (Bertaux et al., 1996; Pérez-Hoyos et al., 2018). Cloud-based organic chemical reactions were also proposed (Spacek et al., 2024), which included possible minor contributions from meteoric carbon. Strikingly, biological molecules (*e.g.*, chlorophylls and iron-sulfur and iron-heme proteins) also overlap with the UV contrasts (Limaye et al., 2018b).

In Venus' sub-cloud atmosphere (<47 km), the visible absorbance (~440–700 nm) was measured by Venera 11, 13, and 14 (Moroz et al., 1979; Ignatiev et al., 1997). Reasonable spectral matches were shown using $S_3$ and $S_4$ (Moroz et al., 1979; Maiorov et al., 2005; Krasnopolsky, 2013). However, the presence of multiple absorbers was also suggested (Maiorov et al., 2005). Chemical models yield support for $S_3$ and $S_4$ below the clouds but yield mixed results for polysulfur species at the cloud tops (Zolotov, 2026). This potentially precludes $S_3$ and $S_4$ from serving as the unified absorbers above and below the clouds. Yet, this opens the possibility that unique absorbers are in the hot sub-cloud atmosphere.

For Venus' cloud aerosols, the UV absorbance (240–400 nm) was inadvertently measured by Vega 1 (Bertaux et al., 1996). During the descent of Vega 1 and 2, the UV spectrometers (ISAV 1 and 2) acquired multiple spectra, with the bulk absorbance being assigned to $SO_2$ (Bertaux et al., 1996). However, at ~ 18 km, the Vega 1 measurements revealed an anomalous absorption feature that was attributed to the unplanned collection of cloud aerosols (Bertaux et al., 1996). After a mechanical shock at ~18 km, these aerosols released from the spectrometer mirrors and into the optical path, thereby yielding the unexpected absorbance spectrum, which was shown to be marginally similar to croconic acid (Bertaux et al., 1996). We recently re-evaluated (Mogul et al., 2025) these observations and proposed that the ISAV spectrometers experienced partial but temporary clogs of the intake inlet due to the inadvertent capture of aerosols beginning at ~52 km, and that the remaining aerosols in the collection tube thermally evaporated and partially decomposed across the descent (~52–0 km, ~330–735 K), thereby yielding the measured UV absorber(s).

In this report, we propose that the UV/blue absorbers from the cloud tops to the sub-cloud atmosphere are derived from cometary organic and inorganic materials via stepwise, altitude-dependent alteration pathways. Cometary dust is rich in organic matter, while oxygen, magnesium, silicon, and iron compounds comprise the major inorganic fraction (Jessberger et al., 1988; Zolensky et al., 2024). Cometary dust likely comprises ~94% (~29 t $d^{-1}$) of the total cosmic input at Venus (Carrillo-Sánchez et al., 2020), while the assessed ablation efficiency is higher than at Mars and Earth (Carrillo-Sánchez et al., 2020). The inferred daily influx of iron at Venus is considerable (~4 t $d^{-1}$) (Carrillo-Sánchez et al., 2020) with a recent microphysics model suggesting substantial delivery of this iron (and magnesium) through the clouds and sub-cloud atmosphere (Karyu et al., 2026). Conversely, the ablation of cosmic organic matter results primarily in conversion to $CO_2$ above the clouds (Carrillo-Sánchez et al., 2020). However, a non-negligible amount of cosmic organic matter survives ablation, resulting in an appreciable yearly influx of organic carbon into Venus' clouds (~6 t $yr^{-1}$) (Carrillo-Sánchez et al., 2020). This 'unablated' carbon includes polycyclic aromatic hydrocarbons (PAHs), insoluble organic matter (IOM), hydrogenated amorphous carbon, and aliphatic hydrocarbons (Sandford et al., 2016; Alexander et al., 2017; Potapov and McCoustra, 2021).

Hence, this work quantitatively and statistically compares Venus' spectra (**Fig. 1**) to mixtures of PAHs and iron-bearing compounds, which serve as compositional analogs for UV-absorbing, altered cosmic dust. We describe the acquisition of possible altitude profiles for the PAHs and iron-bearing compounds and estimate bulk residence times at the cloud tops. We then discuss the proposed unified assignment of altered cometary organics as the primary absorbers across Venus' atmosphere, along with the associated challenges.

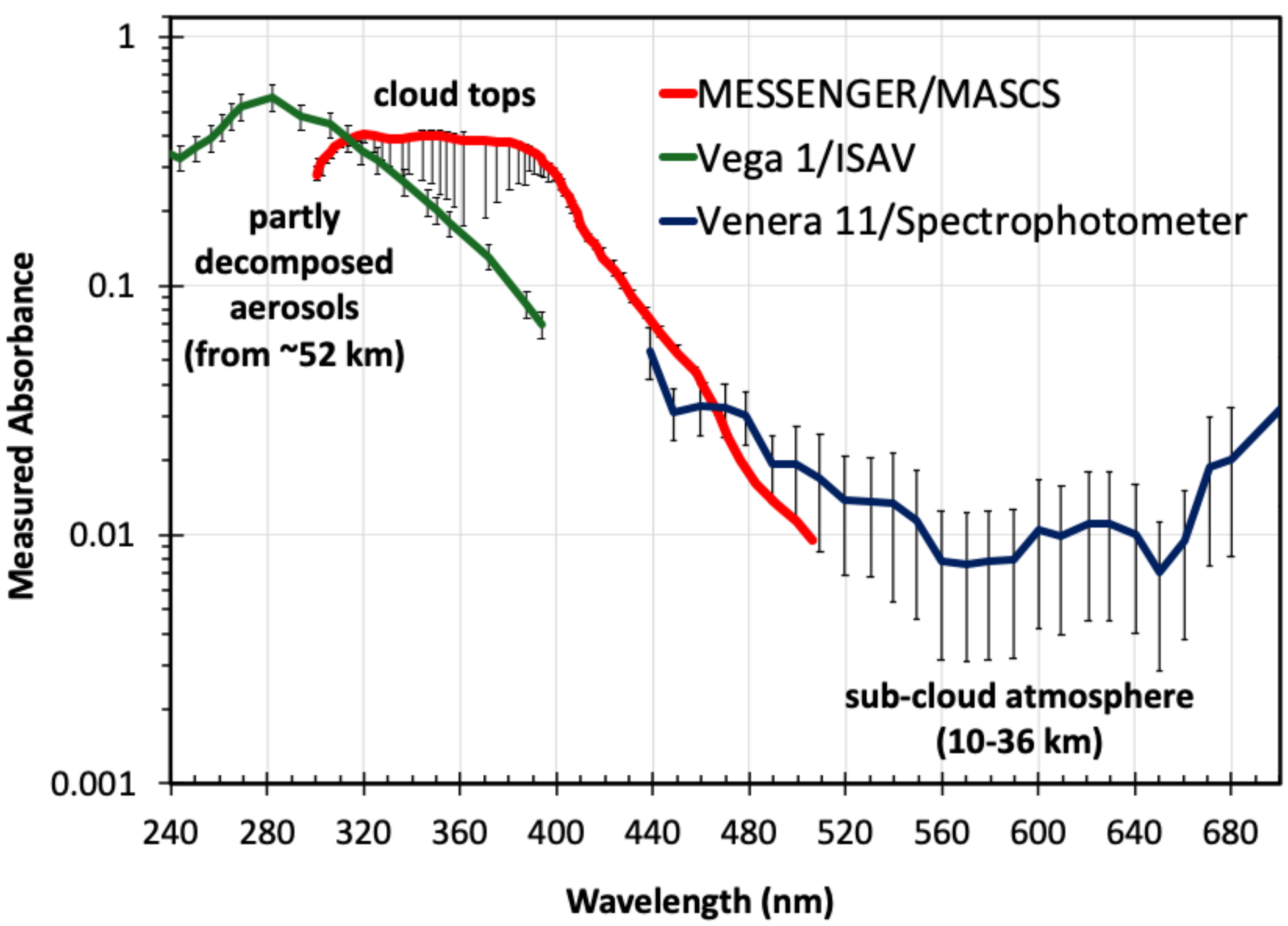

**Fig. 1.** Absorbance spectra for Venus' cloud tops, partly decomposed cloud aerosols, and sub-cloud atmosphere. The measured or true absorbances (log-scale) were calculated as described in the Methods. Cloud top spectrum (75 ± 7 km) was obtained by MESSENGER/MASCS. Spectrum for the partly decomposed aerosols (from ~47–52 km) was obtained by Vega 1 (ISAV 1) at ~18 km. The sub-atmosphere spectrum (10–36 km) was obtained by the Venera 11 spectrophotometer. Error bars represent the reported uncertainties.

## 2. Methods

### *2.1. Sources for the UV/Vis absorption spectra*

#### *2.1.1. Venus' cloud tops, cloud aerosols, and sub-cloud atmosphere*

Digitized spectra were obtained for Venus' cloud tops (Pérez-Hoyos et al., 2018), aerosols (Bertaux et al., 1996), and sub-cloud atmosphere (Maiorov et al., 2005). Altitude profiles for the atmospheric temperatures and pressures at Venus were obtained from Seiff et al. (1985). The absorbance of the cloud tops (300–500 nm), reported as the altitude range of 75 ± 7 km, was measured by the MASCS spectrometer on the MESSENGER probe (Pérez-Hoyos et al., 2018). The extracted MESSENGER spectra included the optimum, maximum, and minimum relative absorbance values from Figure 15 in Pérez-Hoyos et al. (2018) (dashed and solid black lines). The maximum and minimum absorbances represented the upper and lower limits. The relative error for the measurements was reported as 5% (Pérez-Hoyos et al., 2018).

For this study, the MESSENGER relative absorbances were converted to true absorbances (or measured absorbances). To our understanding, the relative absorbances ($A_{rel} = A_{meas}/A_{max}$) were obtained by dividing all measured values ($A_{meas}$) by the maximum measured absorbance ($A_{max}$, ~320 nm). We thus estimated $A_{max}$ by extracting the minimum reflectance value, $(I/F)_{min}$, for the cloud tops from Figure 1 in Pérez-Hoyos et al. (2018) ($(I/F)_{min} = 0.392$ at ~320 nm; light gray plot), followed by conversion to absorbance ($A_{max} = 0.407 = -\log(I/F)_{min}$). The estimated true absorbances were then obtained by multiplying the relative absorbances by $A_{max}$ ($A_{meas} = (A_{rel})(A_{max})$).

For the captured cloud aerosols, the optical depth (240-390 nm) was measured *in situ* by the ISAV 1 spectrometer on Vega 1 (Bertaux et al., 1996). These aerosols were collected between ~52–47 km, while the associated UV spectrum was obtained at 18 km. The optical depth was converted to absorbance ($A = \tau/\ln(10)$). The uncertainties in the ISAV measurements (~ ±12%) were estimated by tracking the scatter in the reported $SO_2$ abundances at 10–40 km (Bertaux et al., 1996), consistent with the suggestion in the caption of Figure 12 in Bertaux et al. (1996).

For the sub-cloud atmosphere, the optical depth (440-650 nm) was measured *in situ* by the Venera 11 spectrophotometer (Maiorov et al., 2005). The spectrum used in this study represented the altitudes of 10–36 km. The optical depth was converted to absorbance. The uncertainties for the spectrum from 10–36 km (*e.g.*, ~41% at 500 nm) were estimated using the relative uncertainties provided for the spectra from 10–19 km and 3–10 km (Maiorov et al., 2005).

*2.1.2. Polycyclic aromatic hydrocarbons*

Calculated cross sections (240–750 nm) for the polycyclic aromatic hydrocarbons (PAHs) were obtained from the Theoretical Spectra Database for Polycyclic Aromatic Hydrocarbons(Malloci et al., 2007) (TSD-PAHs) (https://astrochemistry.oa-cagliari.inaf.it/database/pahs.html). The TSD-PAHs contained 40 spectra for ground-state PAHs (*e.g.*, single molecule in the gas phase). The spectra were obtained using identical methods, conditions, and wavelengths, and possessed sufficient wavelength overlaps with the Venus spectra to permit comparisons. These spectra were independent of the impacts of particle size, film thickness, temperature, or solvents. Absorption cross sections ($cm^2$ $molecule^{-1}$) were converted to molar absorptivity for the data fitting ($M^{-1}cm^{-1} = cm^2\ molecule^{-1} \times \frac{N_A}{mol} \times \frac{1L}{1000\ cm^3}$; $N_A$ = Avogadro constant). Typical absorption cross sections for the 5–10 ring PAHs were $\leq 10^5$ $M^{-1}$ $cm^{-1}$ or $\leq 10^{-16}$ $cm^2$ $molecule^{-1}$ (200–500 nm).

*2.1.3. Iron compounds*

Spectra and cross sections (200–780 nm) for solid-phase rhomboclase ($(H_5O_2)Fe(SO_4)_2(H_2O)_3$), solid-phase acid ferric sulfate ($(H_3O)Fe(SO_4)_2$), and solution-phase rhomboclase (dissolved in 52 wt% $H_2SO_4$) were obtained from Jiang et al. (2024). The respective cross sections ($cm^2$ $g^{-1}$) were converted to $M^{-1}cm^{-1}$ ($M^{-1}cm^{-1} = cm^2 g^{-1} \times MM \times \frac{1L}{1000\ cm^3}$; $MM$ = molecular mass, g/mol). The respective absorption cross sections were ≤200, ≤150, and ≤110 $M^{-1}$ $cm^{-1}$ (200–500 nm).

Spectra and molar absorptivity values (200–500 nm) for solution-phase ferric chloride ($FeCl_3$), ferric sulfate ($Fe_2(SO_4)_3$), and a mixture of ferric chloride and ferric sulfate were obtained from Egan et al. (2025b). The molar absorptivity values for ferric chloride (in 8.96 M HCl) were reported as $M^{-1}cm^{-1}$ (Egan et al., 2025b). The molar absorptivity values for ferric sulfate and the ferric chloride/sulfate mixture were calculated using the reported absorbances and concentrations in Egan et al. (2025b). The concentration of ferric sulfate was reported as 0.34 mM in 12.14 M $H_2SO_4$ (73 wt% $H_2SO_4$) (Egan et al., 2025b). The concentrations for the ferric chloride/sulfate mixture were reported as 82 mol% ferric chloride and 18 mol% ferric sulfate, and 0.172 mM total Fe in 12.8 M $H_2SO_4$ (76.5 wt%) and 0.093 M HCl (Egan et al., 2025b). These concentrations were re-expressed as 0.120 mM ferric chloride and 0.026 mM ferric sulfate. The respective absorption cross sections were ≤5500, ≤5200, and ≤10,000 $M^{-1}$ $cm^{-1}$ (200–500 nm).

The absorbance spectrum for powders of ferric hydroxysulfates, which was described as a mixture of hydronium jarosite ($(H_3O)Fe_3(SO_4)_2(OH)_6$) and volaschioite ($Fe_4(SO_4)O_2(OH)_6 \cdot 2H_2O$), was obtained from El-Hosainy et al. (2021). The absorbance scale for the ferric hydroxysulfates was reported as “K/M” (El-Hosainy et al., 2021), which we interpreted as the Kubelka-Munk transformation for the conversion of reflectance to absorbance for optically thick materials (Myrick et al., 2011). The molar absorptivity values for the ferric hydroxysulfate mixture were thus estimated using the apparent absorbances (El-Hosainy et al., 2021) and total concentration of the ferric hydroxysulfates (~21 mM). The concentration of ferric hydroxysulfates was calculated using the initial conditions of the sulfate-forming reaction (El-Hosainy et al., 2021) (4.45 mmoles of $Fe(NO_3)_3(H_2O)_9$ starting material, and 61.92 mL total volume) and assuming the formation of a 1:1 mole mixture of hydronium jarosite and volaschioite. The inferred absorption cross sections were ≤170 $M^{-1}$ $cm^{-1}$ (200–500 nm).

## *2.2. Spectral fits*

### *2.2.1. Selection of PAHs and iron compounds*

Candidate PAHs from the TSD-PAHs (Malloci et al., 2007) were chosen based on the degree of overlap with the Venus spectra at the major wavelengths of UV and/or visible absorption. For the cloud top spectra, candidate PAHs exhibited notable absorbances at 300–400 nm, which was the wavelength range where the MESSENGER/MASCS instrument measured the maximum absorbances. These PAHs collectively possessed 4–24 rings and included triphenylene ($C_{18}H_{12}$; 4 rings), coronene ($C_{22}H_{12}$; 7 rings), pentacene ($C_{22}H_{14}$; 5 rings), hexacene ($C_{26}H_{16}$; 6 rings), ovalene ($C_{32}H_{14}$; 10 rings), and circumovalene ($C_{66}H_{20}$; 24 rings).

For the captured cloud aerosol spectrum (from 18 km), candidate PAHs exhibited notable absorbances at 270–290 nm, which was the wavelength range where the ISAV 1 instrument measured the maximum absorbances. These PAHs collectively possessed 3–4 rings and included anthracene ($C_{14}H_{10}$; 3 rings), chrysene ($C_{18}H_{12}$; 3 rings), phenanthrene ($C_{14}H_{10}$; 3 rings), pyrene ($C_{16}H_{10}$; 4 rings), tetracene ($C_{18}H_{12}$; 4 rings), and triphenylene ($C_{18}H_{12}$; 4 rings).

For the sub-cloud atmosphere spectrum (10–36 km), candidate PAHs exhibited notable absorbances at 400–550 nm, which was the wavelength range where the Venera 11 spectrophotometer measured the maximum absorbances. Two sets of candidate PAHs were chosen. The first set was identical to the composition of the cloud top PAHs. The second set contained a unique set of PAHs. This second set possessed 5–19 rings and included circumcoronene ($C_{54}H_{18}$; 19 rings), circumpyrene ($C_{42}H_{16}$; 14 rings), dibenzo[bc,ef]coronene ($C_{30}H_{14}$; 9 rings), hexabenzocoronene ($C_{42}H_{18}$; 13 rings), and perylene ($C_{20}H_{12}$; 5 rings).

The tested iron compounds included solution-phase ferric chloride (in ~32 wt% HCl), solution-phase mixtures of ferric chloride and sulfate (in ~77 wt% $H_2SO_4$ with minor amounts of HCl), solution-phase ferric sulfate (in ~73 wt% $H_2SO_4$), solution-phase or dissolved rhomboclase (in ~52 wt% $H_2SO_4$), solid-phase rhomboclase, solid-phase acid ferric sulfate, and powders of ferric hydroxy sulfates.

*2.2.2. Multi-component spectral fitting and statistical treatments*

The absorbance spectra for Venus' cloud tops, captured cloud aerosols, and sub-cloud atmosphere were independently fitted using the multi-component mixtures described above. Fits were conducted with mixtures containing ≤6 PAHs and ≤2 iron compounds, which included fits with 1 PAH or 1 iron compound. Thus, the total number of fitted parameters ($k$) ranged from 1 to 7. Through **Equation 1**, the total absorbance ($A_T$) was obtained by summing the contributions from each component ($i$) in the system ($k$) at each wavelength ($\lambda$). The absorbance values for each species were calculated (**Equation 1**) via the Beer-Lambert law ($A = \varepsilon lc$) using the described molar absorptivity values ($\varepsilon_i$) at each wavelength, the appropriate/approximate pathlength ($l_i$), and the respective concentration terms ($c_i$). In the data fitting, the concentration terms were the unknown variables.

$$A_T(\lambda) = \sum_i^k A_i(\lambda) = \sum_i^k \varepsilon_i(\lambda) l_i c_i \qquad \textbf{(1)}$$

The pathlength for the MESSENGER/MASCS spectra was set to 14 km, which represented the reported altitude range for UV absorption (75 ± 7 km) (Pérez-Hoyos et al., 2018). The reported optical pathlength of 170 cm was used for the Vega 1 spectrum (Bertaux et al., 1996). A nominal pathlength of 26 km, or the difference between the altitudes of 10 to 36 km, was used for the sub-cloud atmosphere spectrum from Venera 11.

Regressions were conducted by summing and minimizing the chi-squared statistic ($\chi^2$) using **Equation 2**, where $y_n$ was the observed measurement, $f(x_n)$ was the calculated value from **Equation 1**, $\sigma$ was the uncertainty or error in the spectral measurement, and $N$ was the total number of fitted data points. For the cloud tops (300–450 nm), the relative error of the measurement was 5%, as reported in Pérez-Hoyos et al. (2018). For the aerosols and sub-cloud atmosphere, the respective measurement uncertainties were ~12 and 41% (as detailed above), which were potentially overestimated. Spectral fits were independently ranked for the cloud top maximum absorbance ($\chi_1^2$), cloud top minimum absorbance ($\chi_2^2$), captured cloud aerosol absorbance ($\chi_3^2$), and sub-cloud atmosphere absorbance ($\chi_4^2$).

$$\chi^2 = \sum_{n=1}^{N} \left( \frac{(y_n - f(x_n))}{\sigma} \right)^2 \quad \textbf{(2)}$$

$$\chi_{red}^2 = \chi^2 / \nu \quad \textbf{(3)}$$

$$\mathrm{AIC} \approx \chi^2 + 2k \quad \textbf{(4)}$$

Goodness of the spectral fits and plausibility of the regression models were assessed using the reduced chi square statistic ($\chi_{red}^2$) and the Akaiki Information Criterion (AIC) (Liddle, 2007; Landay et al., 2017). The $\chi_{red}^2$ term (Livadiotis, 2025) was obtained from **Equation 3** using the degrees of freedom ($\nu = N - k$) and number of fitted parameters ($k \leq 7$) in the initial regression conditions. The AIC term (Landay et al., 2017) was estimated through **Equation 4** and the ΔAIC calculated against the lowest value. Good fits were inferred when $\chi_{red}^2 \approx 1$ (Livadiotis, 2025). However, for the Vega 1 and Venera 11 spectra, this was interpreted as $\chi_{red}^2 \approx 0.2 - 1$ to roughly correct for the impacts of the high measurement uncertainties, which artificially skew the $\chi_{red}^2$ value downwards (Livadiotis, 2025). Support for the regression model was inferred by minimized AIC values. Models were inferred as indistinguishable when ΔAIC ≤ 2, very similar when ΔAIC = 3, potentially comparable when ΔAIC = 4–7, marginally comparable when ΔAIC = 8–9, and likely not comparable when ΔAIC ≥ 10 (Burnham and Anderson, 2004). Overfitting was inferred when increasing the number of fit parameters increased the AIC.

### *2.3. Calculating concentrations*

Using the best spectral fits and true absorbances, the concentrations (nM or pM) for the individual PAHs, total PAHs, and iron compound/s were obtained for the cloud tops, aerosols, and sub-cloud atmosphere. The concentrations of PAHs and iron compounds were converted to total organic carbon and iron using the corresponding molecular formulas. For the cloud tops, the concentrations were calculated as the averaged values from the fitted maximum and minimum absorbance spectra, with the uncertainties representing the total ranges between the maximum and minimum concentrations. For the aerosol and sub-cloud concentrations, the uncertainties from the Vega 1 and Venera 11 spectra were propagated.

For the captured and partly decomposed aerosols, the *apparent* concentrations were normalized to roughly account for the yields of aerosol capture, thermal decomposition, and release from the mirrors in the Vega 1 collection tube. **Equation 5** yielded the estimated aerosol PAH concentrations ($[PAH]_{aerosol}$) using a normalization factor ($[Fe]_{aerosol}/[Fe]_{Vega}$), which was defined as the ratio between the reference aerosol Fe concentration ($[Fe]_{aerosol}$) and the *apparent* Fe concentration measured by Vega 1 ($[Fe]_{Vega}$). The LNMS value for aerosol Fe concentration (Mogul et al., 2026) was assigned as the reference value ($[Fe]_{aerosol} = [Fe]_{LNMS} = 16 \pm 9$ nM or $0.9 \pm 0.5$ mg m$^{-3}$). Thus, the $[PAH]_{aerosol}$ values were obtained by multiplying the *apparent* PAH concentrations ($[PAH]_{Vega}$) by the normalization factor ($[Fe]_{LNMS}/[Fe]_{Vega} = 2.5\times10^{-3} \pm 1.4\times10^{-3}$).

$$[PAH]_{aerosol} = [PAH]_{Vega} \times \frac{[Fe]_{LNMS}}{[Fe]_{Vega}} = [PAH]_{Vega} \times \frac{[Fe]_{aerosol}}{[Fe]_{Vega}} \qquad \textbf{(5)}$$

***2.4 Calculating bulk residence times***

Bulk residence times ($\tau$) for the PAHs and iron compounds at the cloud tops were obtained using **Equation 6**. Bulk residence times were treated as the accumulation time, or the time required to accumulate the target concentration via cosmic dust influx. Relevant terms included the maximum concentrations for the total organic carbon or iron from the fitted cloud top spectra ($M$, expressed as mol m$^{-3}$), shell volume between 75±7 km ($V = 6.6\times10^{18}$ m$^3$), respective molar mass (g/mol), and infall rate for the cosmic materials ($\phi$, t yr$^{-1}$). The infall rate at Venus for 'unablated carbon' was $0.017 \pm 0.009$ t d$^{-1}$ ($6.2 \pm 3.6$ t yr$^{-1}$) (Carrillo-Sánchez et al., 2020), which included organic carbon that survived ablation, which was assumed to be predominantly composed of aromatic organics for this study. The infall rate at Venus for 'ablated iron' was $4.1 \pm 2.4$ t d$^{-1}$ (1500 $\pm$ 900 t yr$^{-1}$) (Carrillo-Sánchez et al., 2020), which included the iron available after ablation to yield ferric chloride and/or sulfates. Uncertainties for the infall rates were proportionately assigned using the total cosmic influx rate (31 ± 18 t d$^{-1}$) (Carrillo-Sánchez et al., 2020). All uncertainties were propagated.

$$\tau = \frac{MV(g/mol)}{\phi} \quad (6)$$

## 3. Results

**Table 1.** List of candidate polycyclic aromatic hydrocarbons (PAHs) and iron compounds and their abbreviations.

| PAHs | Abbreviation | PAHs | Abbreviation | Iron Compounds (physical state) | Abbreviation |
|---|---|---|---|---|---|
| **Anthracene** | Ant | **Hexacene** | Hex | **Ferric Chloride** (aqueous, in HCl) | FC |
| **Chrysene** | Chr | **Pentacene** | Pen | **Ferric Chloride and Ferric Sulfate** (aqueous, in $H_2SO_4$ and HCl) | FC/FS |
| **Coronene** | Cor | **Perylene** | Per | **Ferric Sulfate** (aqueous, in $H_2SO_4$) | FS |
| **Circumcoronene** | Cic | **Phenanthrene** | Phe | **Rhomboclase** (aqueous, in $H_2SO_4$) | sRho |
| **Circumpyrene** | Cip | **Pyrene** | Pyr | **Rhomboclase** (solid phase) | Rho |
| **Circumovalene** | Cio | **Ovalene** | Ova | **Acid Ferric Sulfate** (solid phase) | AFS |
| **Dibenzo[bc,ef]coronene** | Dbc | **Tetracene** | Tet | **Ferric Hydroxysulfates** (hydronium jarosite and volaschioite; solid phase) | FHS |
| **Hexabenzocoronene** | Hbc | **Triphenylene** | Tri | | |

### *3.1. Multi-component spectral analysis*

The absorbance spectra (**Fig. 1**) for Venus' cloud tops (Pérez-Hoyos et al., 2018), captured cloud aerosols (Bertaux et al., 1996), and sub-cloud atmosphere (Maiorov et al., 2005) were independently fitted using multi-component mixtures containing different combinations of PAHs and iron compounds. These mixtures align with the heterogeneity of altered cosmic dust (Okumura and Mimura, 2011; Sephton et al., 2024; Aponte et al., 2026). Unique mixtures yielded very good matches to the respective Venus spectra (**Figs. 2-4**). **Table 1** provides the candidates and abbreviations for the PAHs (possessing 3–24 rings) and iron-bearing compounds used in this study (see **Section 2.2.1**). **Appendix A** provides statistical support for the multi-component absorption model. **Appendix B** explains how the spectral fits were interpreted and describes the knowledge gaps and uncertainties associated with developing a radiative transfer model for a mixture of PAHs and iron. **Appendix C** provides an assessment of the expected physical states in the cloud tops, decomposed cloud aerosols, and sub-cloud atmosphere.

For the cloud top absorbances, the spectra (300–500 nm) representing the maximum and minimum absorbances were from the MESSENGER/MASCS instrument (Pérez-Hoyos et al., 2018). The spectrum (230–390 nm) for the captured and partly decomposed cloud aerosols was from Vega 1 at 18 km (Bertaux et al., 1996). The sub-cloud atmosphere spectrum (440–1200 nm) was from Venera 11 at 10–36 km (Moroz et al., 1979).

The Venus spectra were fitted using mixtures containing ≤6 PAHs and ≤2 iron compounds (**Appendix Figs. D.1-4**). In all cases, the choice of 6 PAHs was excessive, as the regression steps eliminated several PAHs. Goodness of the spectral fits were assessed using the chi squared statistic ($\chi^2$), which was calculated using the errors/uncertainties of the MESSENGER (5%), Vega 1 (12%), and Venera 11 (41%) measurements (see **Section 2.1.1**). The residuals from the best fits showed no systematic trends and were generally consistent with the respective errors/uncertainties of the Venus spectra. Qualities of the regression models (*e.g.*, composition and number of fitted parameters) were inferred using the reduced chi square statistic ($\chi^2_{red}$) and AIC terms (see **Section 2.2.2**), where the AIC included a penalty for increasing complexity of the model. Good fits to the data were inferred when the $\chi^2_{red} \approx 1$. A plausible model was inferred when the AIC value was minimized. Similarities between tested models were inferred when $\Delta$AIC ≈ ≤3. The inclusion of multiple absorbers (PAHs and iron) was supported by the $\chi^2_{red}$ and $\Delta$AIC terms, as described in **Appendix A**. The $\chi^2$ for selected spectral fits are listed in **Appendix Figs. D.1-4**. The $\chi^2_{red}$ and AIC terms are compared in **Appendix Figs. D.5-7**.

*3.2. Outcomes of the spectral fits*

For the cloud tops (300–500 nm), the best spectral matches were obtained with a mixture of 3 PAHs (5–10 rings) + FC (**Figs. 2** and **Appendix Fig. D.1a**), where the PAH composition was: ovalene > coronene ≈ pentacene. Among the tested combinations, this mixture (3 PAHs + FC) yielded an $\chi^2_{red} \approx 1$ and a minimized AIC value; thereby suggesting a good fit and a plausible multi-component model (**Appendix Figs. D.5a** and **D.6a**). For the maximum absorbance spectrum, very good and virtually identical fits were obtained with PAHs + FC/FS or just PAHs ($\chi^2_{red} \approx 1$; ΔAIC ≈ 3-5) (**Appendix Figs. D.1b, D.1c, D.5a, D.6a)**. For the minimum absorbance, moderate fits were obtained with PAHs + FC, just PAHs, or PAHs + FC/FS ($\chi^2_{red} \approx 5$ and ΔAIC ≈ 8-9) (**Appendix Figs. D.2a-c**, **5b**, and **D.6b**). For the maximum and minimum absorbances, relatively poor fits were obtained with just FC, especially at 300–350 nm ($\chi^2_{red} >> 1$; ΔAIC ≈ 241–318) (**Appendix Figs. D.1d**, **D.2d**, **D.5a**, **D.5b**, **D.6a**, and **D.6b**). Poor fits were obtained when using 1 or 2 PAHs ($\chi^2_{red} >>> 1$; ΔAIC ≈ 423–5071) (**Appendix Figs. D.5a**, **D.6a**, and **D.7a**).

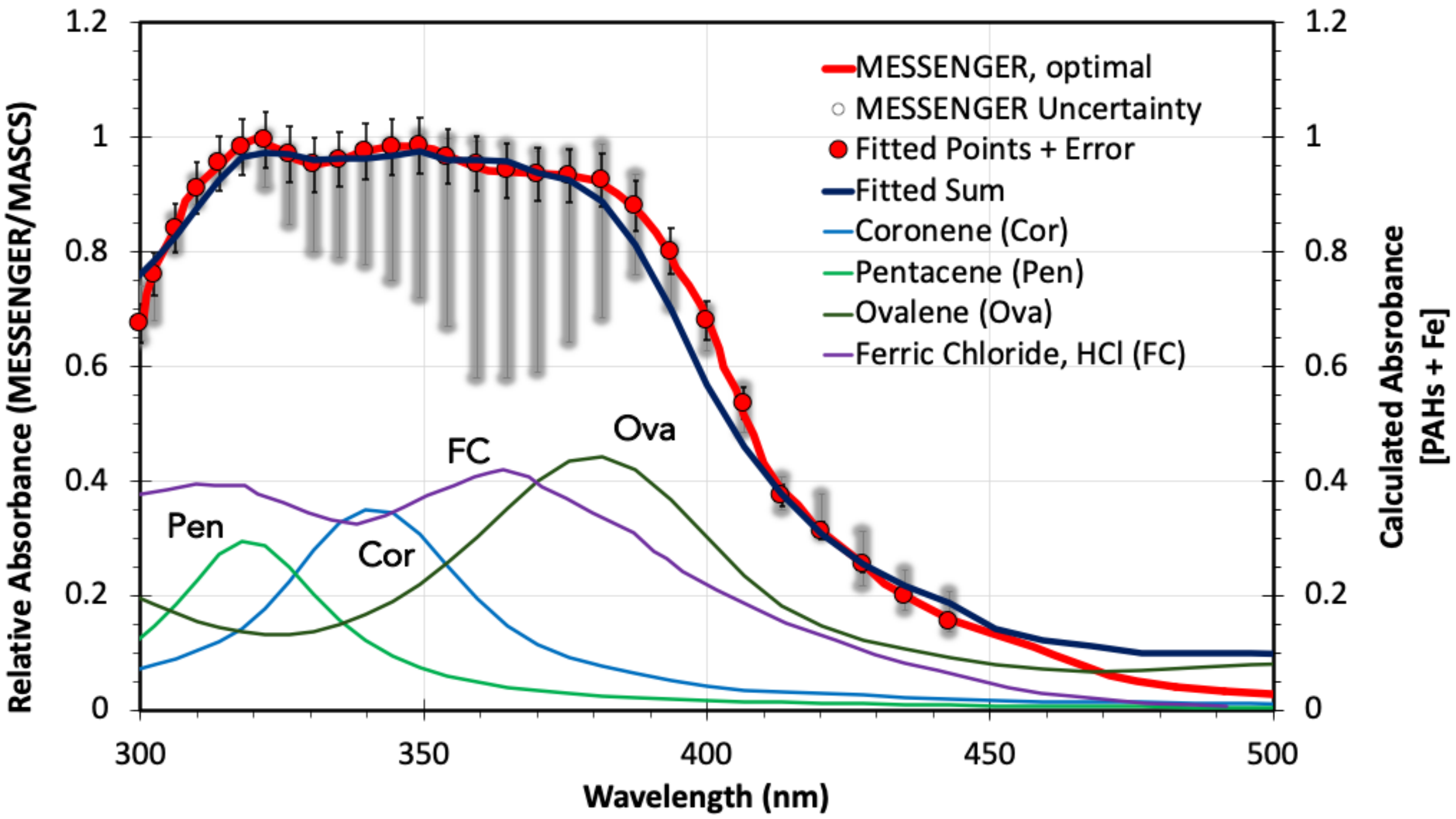

**Fig. 2.** Spectral fit to the relative optimal absorbance for the cloud tops. The absorbance spectrum (75 ± 7 km) was obtained by MESSENGER/MASCS (left-hand y-axis) and fitted with a mixture of PAHs and iron compounds (right-hand y-axis). The composition is summarized in the plot legend and detailed in **Appendix Table D.1**. The red line is the cloud top absorbance, red circles are the fitted data points, and the blue line is the summed absorbance of the mixture. Gray error bars are the reported maximum and minimum uncertainty limits; black error bars represent the relative error reported for the measurements (5%).

For the captured and partly decomposed cloud aerosols, near-perfect matches (240-390 nm) were obtained with a mixture of PAHs (3–4 rings) + AFS (**Fig. 3**, **Appendix Fig. D.3a**), where the PAH composition was: triphenylene > anthracene > chrysene > tetracene. Among the tested combinations, this mixture (PAHs + AFS) yielded an $\chi^2_{red} \approx 1$ (when accounting for the higher uncertainties of the Vega 1 data) and a minimized AIC value, indicating a good fit and plausible model (**Appendix Figs. D.5c** and **D.6c**). Compared to the cloud tops, the Vega 1 spectrum exhibited shorter wavelengths of maximum absorption, indicating a unique molecular composition (**Fig. 1**). This aligns with the presence of PAHs with lower ring numbers. The 3–4 ring PAHs yielded excellent matches at 230–300 nm, while the inclusion of AFS yielded excellent matches at 300–380 nm. Reasonable fits ($\chi^2_{red} \approx$ 2–3) with low support (ΔAIC ≈ 29-35) were obtained with 4 PAHs + FHS and 4 PAHs (**Appendix Figs. D.3b**, **D.3c**, **D.5c**, and **D.6c**). Moderate fits ($\chi^2_{red} \approx$ 5–9) with limited support (ΔAIC ≈ 60–140) were obtained with PAHs + FC/FS, PAHs + FC, PAHs + FS, PAHs + Rho, and PAHs (**Appendix Figs. D.3d-h**, **D.5c**, and **D.6c**). A relatively poor fit was obtained with just FS ($\chi^2_{red} >> 1$; ΔAIC ≈ 200), especially at 230–260 nm (**Appendix Fig. D.3i**). Poor fits were obtained with just 1 PAH (chrysene or tetracene) ($\chi^2_{red} >>> 1$; ΔAIC ≈ 140–500) (**Appendix Figs. D.5c**, **D.6c**, and **D.7b**).

For the sub–cloud atmosphere spectrum obtained by Venera 11, an overlay against the MESSENGER spectra (**Fig. 1**) revealed an overlap (within error) in the apparent true absorbances at ~440–510 nm. This suggested a continuum of absorption across the UV and visible wavelengths between the cloud top and sub-cloud spectra. Alternatively, the Venera 11 spectrum potentially represented a unique composition. To account for both possibilities, two sets of candidate PAHs were chosen for the data fitting. The first set was identical to the cloud top PAH composition (5–10 rings), which assumed no change in PAH identity across the atmosphere. The second set contained unique PAHs (9–13 rings), which assumed that PAH identities would change across the atmosphere. Very good matches were obtained using (listed in order of increasing $\chi^2$): 3 PAHs (9–13 rings) + FHS, 2 PAHs (5–10 rings) + FHS, and 1 PAH (7 rings).

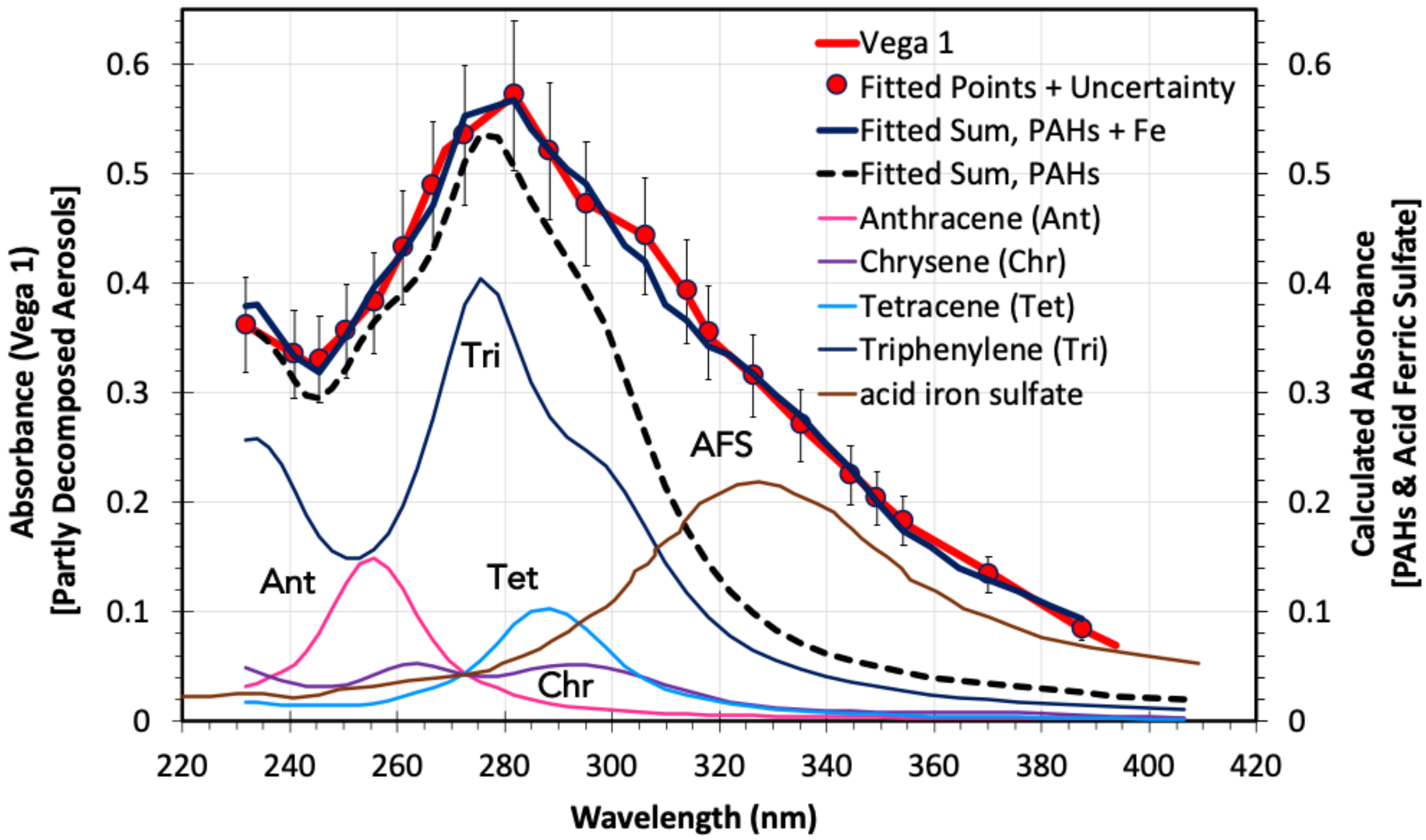

**Fig. 3.** Spectral fit to the absorbance of the partly decomposed aerosols. Aerosols were collected beginning at ~52 km and the absorbance spectrum was measured at 18 km by the ISAV 1 instrument on Vega 1 (left-hand y-axis) and fitted with a mixture of PAHs and iron compounds (right-hand y-axis). The composition is summarized in the plot legend and detailed in **Appendix Table D.1**. The red line is the aerosol absorbance, red circles are the fitted data points, and the blue line is the summed absorbance of the mixture. Black error bars represent the relative error reported for the measurements (12%).

For the spectral fit with 9–13 ring PAHs + FHS (**Appendix Fig. D.4a**), the PAH composition was: dibenzo[bc,ef]coronene > hexabenzocoronene > perylene. For the spectral fit with 5–10 ring PAHs + FHS (**Fig. 4**, **Appendix Fig. D.4b**), the PAH composition was: coronene ≈ pentacene. For the spectral fit with 1 PAH (and no iron), the composition included only coronene (**Appendix Fig. D.4c**); however, pentacene was included as an initial fit parameter but was minimized to zero through the regression. Across these compositions, the AIC value was minimized using 1 PAH, while slightly higher AIC values were obtained for 2 PAHs + FHS (ΔAIC ≈ 3) and 3 PAHs + FHS (ΔAIC ≈ 8) (**Appendix Figs. D.5d** and **D.6d**). Given these ΔAIC values, the models were treated as plausible (2 PAHs + FHS) and marginally plausible (3 PAHs + FHS). Incomplete fits were obtained using just FHS due to a lack of absorbance above ~580 nm (**Appendix Fig. D.4d**)

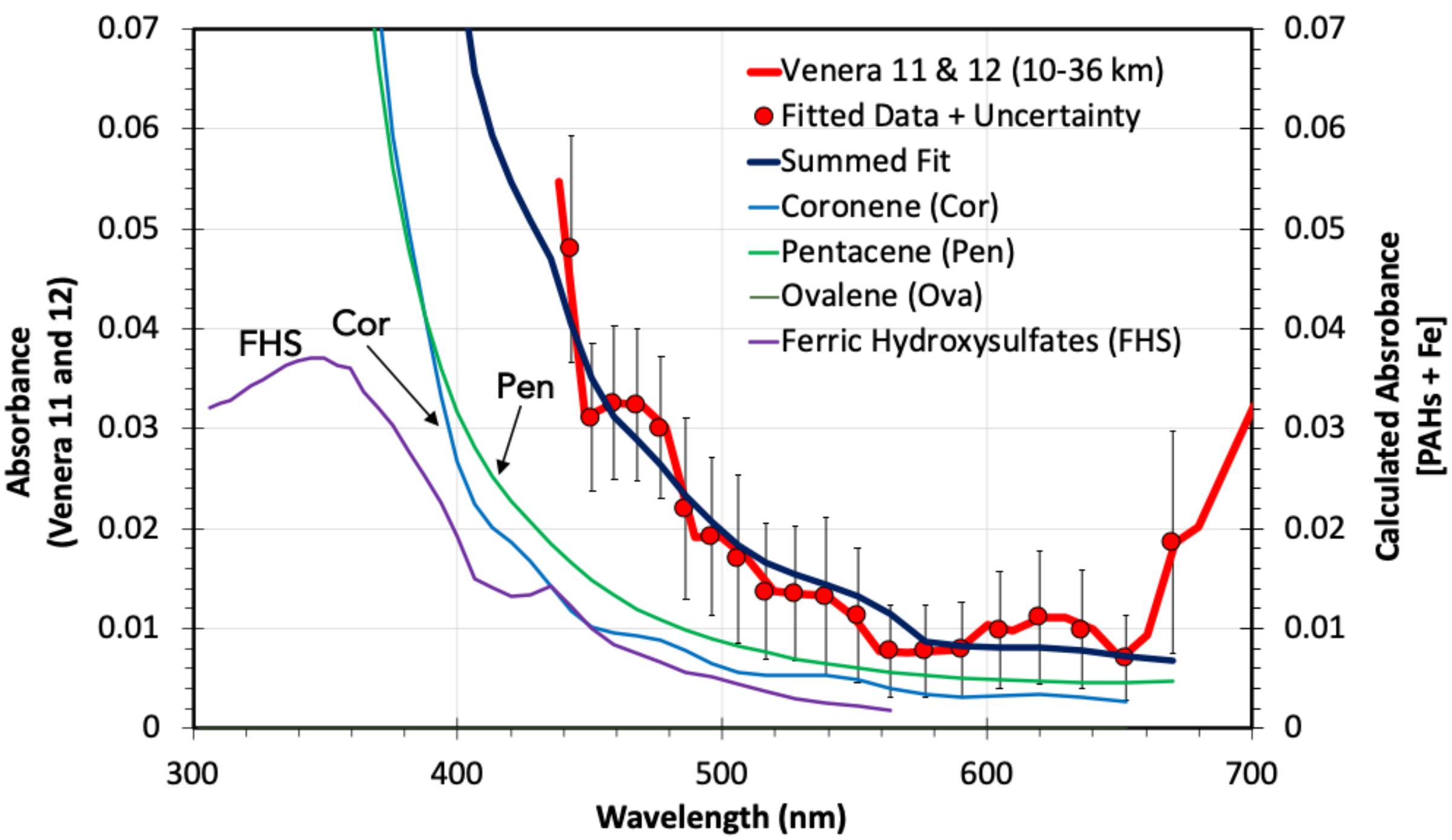

**Fig. 4.** Spectral fit to the sub-cloud atmosphere absorbance. The absorbance spectrum was measured at 10-36 km by the spectrophotometer on Venera 11 (left-hand y-axis) and fitted with a mixture of PAHs and iron compounds (right-hand y-axis). The composition is summarized in the plot legend and detailed in **Appendix Table D.1**. The red line is the absorbance for the sub-cloud atmosphere, red circles are the fitted data points, and the blue line is the summed absorbance of the mixture. Black error bars represent the relative error reported for the measurements (~41%).

### *3.3. Concentrations, altitude profiles, and residence times*

**Table 2.** Molecular and atomic compositions that best fit Venus' UV and blue spectra. Normalized values for the partly decomposed aerosols are provided.

| Atmosphere | | Molecular Composition | | Atomic Composition | |
|---|---|---|---|---|---|
| **Spectra** | **Altitudes (km)** | **Total PAHs (pM)** | **Iron Compound (pM)** | **Total C (PAHs) (pM)** | **Total Fe (pM)** |
| **Cloud Tops** | 75 ± 7 | 3.4 ± 0.3 (5–10 rings) | 15 ± 11 (Ferric Chloride) | 97 ± 8 | 15 ± 11 |
| **Partly Decomposed Aerosols (*normalized*)** | ~52–47 | 94 ± 52 (3–4 rings) | 16,000 ± 12,000 (Acid Ferric Sulfate) | 1600 ± 900 | 16,000 ± 12,000 |
| **Sub-Cloud Atmosphere** | 36–10 | 1.7 ± 0.5 (5–10 rings) | 83 ± 34 (Ferric Hydroxysulfates) | 38 ± 16 | 290 ± 120 |

**Table 2** lists the concentrations for the *total* PAHs, iron compounds, aromatic carbon, and iron from the best spectral fits. **Appendix Table D.1** lists the concentrations for the individual PAHs. As described in **Appendix B**, the *total* concentrations for the PAHs and iron were interpreted as lower limits, while the changes in PAH composition across the altitude profiles were interpreted changes in the general properties of the PAHs (*e.g.*, ring numbers). These treatments roughly accounted for the knowledge gaps pertaining to the physical states, particle sizes, and spectra of the PAHs and iron-bearing compounds.

At the cloud tops, the total PAH concentration was ~4-fold lower than ferric chloride, while the total carbon concentration was ~6-fold greater than total iron. In the partly decomposed aerosols, which were normalized to the LNMS aerosol iron abundance, the total PAH concentration was ~170-fold lower than acid ferric sulfate, while the total carbon concentration was ~10-fold lower than total iron. For the sub-cloud atmosphere, when considering the 2 PAHs (5–10 rings) + FHS mixture, the total PAH concentration was ~50-fold lower than the ferric hyrdoxysulfates, while the total carbon concentration was respectively ~8-fold lower than total iron.

For the cloud tops and sub-cloud spectra, the spectral fits with similar PAHs (coronene, pentacene, and ovalene) showed no changes in coronene and pentacene concentrations (within error); whereas ovalene, the most abundant PAH at the cloud tops, was depleted in the sub-cloud atmosphere (**Fig. 5a**). From the cloud tops to the cloud aerosols (from ~48-51 km), the total concentrations of PAHs and iron compounds (**Fig. 5b**) increased by factors of 27 ± 15 and ~1100 (or range of ~150–7800-fold, considering the uncertainty). The associated total carbon and iron concentrations increased by 17 ± 9 and ~1100 (~150–7800-fold) (**Fig. 5c**). From the cloud aerosols to the sub-cloud atmosphere, the PAHs and iron compounds decreased by factors of 54 ± 19 and ~190 (range of ~33–570-fold) (**Fig. 5b**). The associated total carbon and iron concentrations decreased by 43 ± 27 and ~55 (range of ~9–160-fold) (**Fig. 5c**). Bulk residence times, or accumulation times, at the cloud tops for the total organic carbon (1.3 ± 0.7 Myr) and iron (6.0 ± 3.5 kyr) were estimated using the assessed influx rates of unablated organic carbon and ablated iron, and the reported column of UV absorbance (75 ± 7 km) (Pérez-Hoyos et al., 2018).

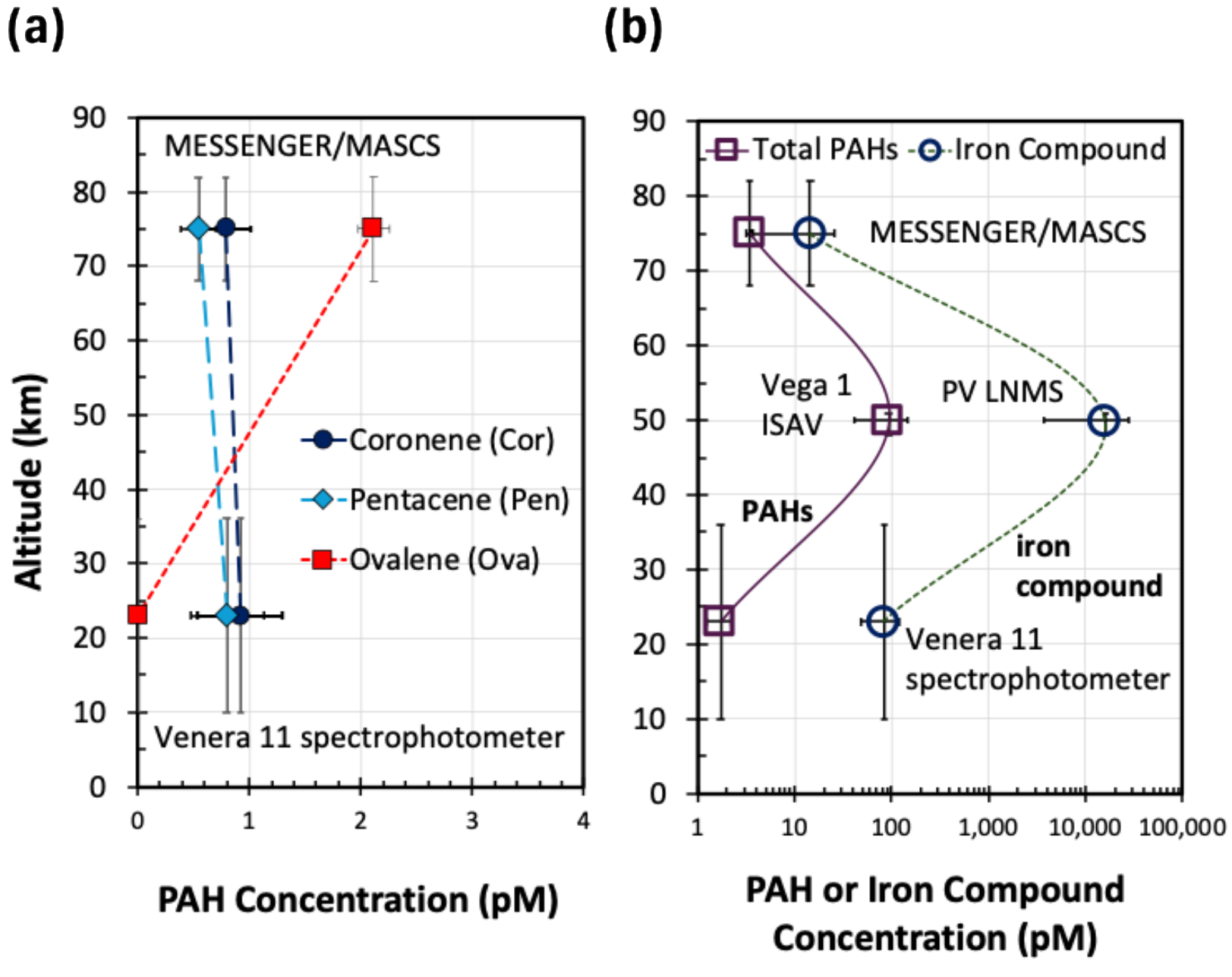

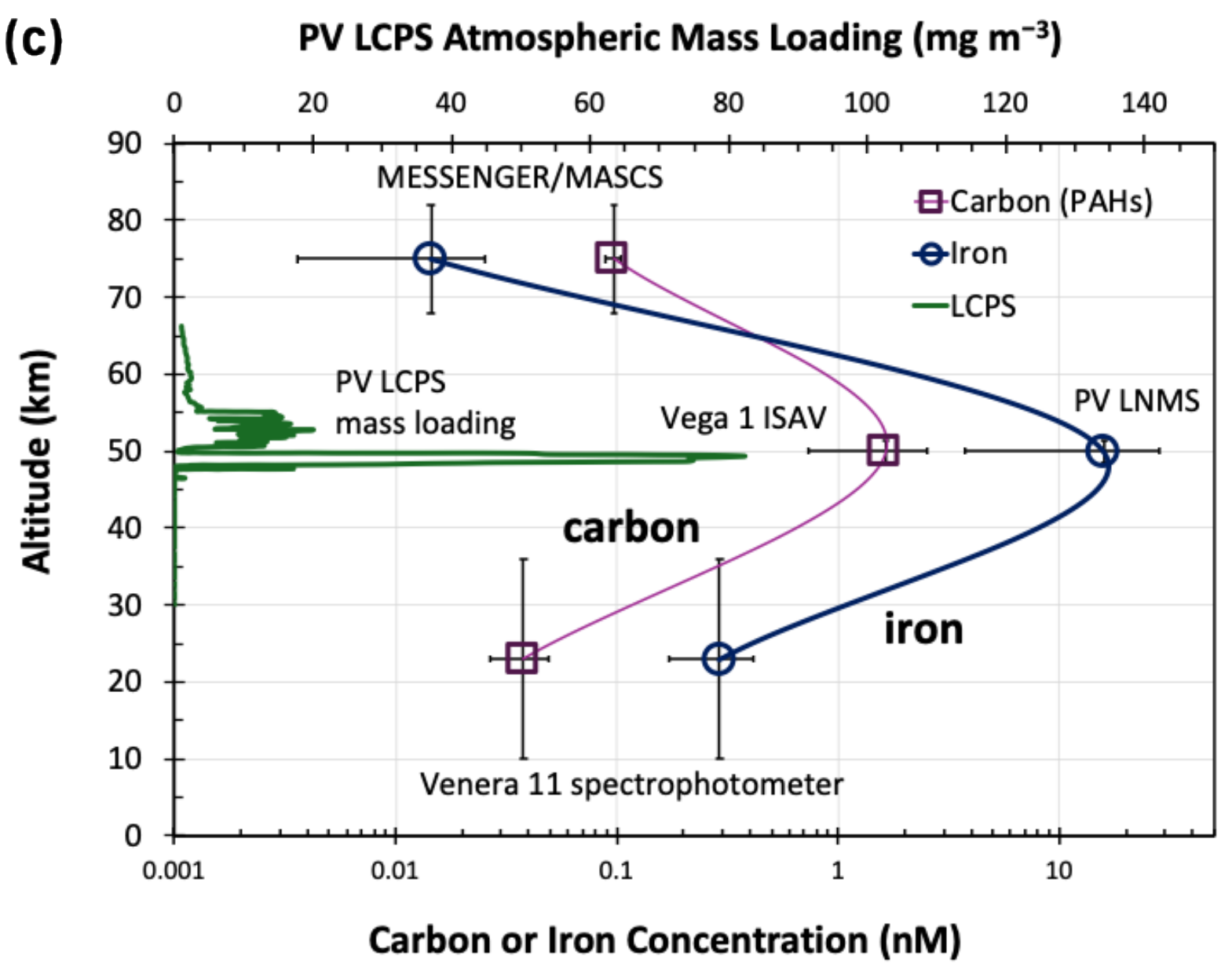

**Fig. 5.** Altitude profiles for (a) individual PAHs (excluding the decomposed cloud aerosols), (b) total PAHs and iron compounds, and (c) total aromatic carbon and iron. Values for the cloud tops, partly decomposed aerosols, and sub-cloud atmosphere were derived by fitting the MESSENGER (75 ± 14 km), Vega 1 (~50 ± 1.5 km), and Venera 11 (23 ± 13 km) spectra. Concentrations (bottom x-axes) were obtained using the true/measured absorbances, as described in the Methods. For Vega 1, values for the captured aerosols were normalized to the PV LNMS value for aerosol iron concentration. In (c), the atmospheric mass loading values (upper x-axis, mg $m^{-3}$) from the PV Large Probe Cloud Particle Size Spectrometer (LCPS) (Knollenberg and Hunten, 1980) are compared to the carbon and iron concentrations.

## 4. Discussion

We demonstrate that mixtures of PAHs and iron compounds yield excellent matches to Venus' spectra from different altitudes across the UV and visible wavelengths (**Fig. 1**). Unique molecular compositions (**Figs. 2-4**) match the cloud tops (5–10 ring PAHs and ferric chloride), partly decomposed cloud aerosols (3–4 ring PAHs and acid ferric sulfate), and sub-cloud atmosphere (5–10 ring PAHs and ferric hydroxy sulfates). Several knowledge gaps, as detailed in **Appendix B**, present challenges towards developing a low-uncertainty radiative transfer model for mixtures of PAHs, iron, and sulfuric acid. To account for these knowledge gaps, and in the absence of a robust radiative transfer model, the peak fitting results were interpreted as proxies for the bulk changes in the total PAH and carbon concentrations and general properties of the PAHs (*e.g.*, ring numbers, reactivity in $H_2SO_4$, thermal stability, and red shifts in the absorbance); where the total PAH and iron concentrations were interpreted as lower limits.

Under this interpretation, the PAH composition (5–10 rings) at the cloud tops (75 ± 7 km; Pérez-Hoyos et al. (2018)) aligns with the release of PAHs into the clouds by unablated organic matter, formation of PAHs through partial thermal alteration of organic matter, and/or formation of PAHs through impact synthesis at or above the cloud tops (Clemett et al., 2010). For example, 2–6 ring PAHs are constituents of unablated carbonaceous chondrites (Clemett et al., 2010). Laboratory heating (553 K) of CI-type chondritic materials, such as the returned Bennu sample, yields gases containing 2-6-ring PAHs (Aponte et al., 2026). The pyrolysis of chondritic organic matter (at ~550–920 K) (Okumura and Mimura, 2011; Sephton et al., 2024), coal tar (at ~1280 K) (Ledesma et al., 2000), and benzene (at ~1100 K) (Marsh et al., 2005) yields the release of 3–10 ring PAHs. The UV irradiation of hydrogenated amorphous carbon may also yield PAHs (Scott et al., 1997).

For the captured aerosols (from ~47–52 km), the PAH composition (3–4 rings) aligns with the conditions of the proposed Vega 1 descent sequence (Mogul et al., 2025). This includes the capture of aerosol-phase organics at ~52 km (*e.g.*, 5–10 ring PAHs and IOM), partial thermal decomposition of the aerosol contents from ~52–18 km, and the release of PAHs with lower ring numbers (and other materials) from the ISAV surfaces at ~18 km. In support, 3–4 ring PAHs form during the pyrolysis of chondritic organic matter beginning at ~600 K (Okumura and Mimura, 2011; Sephton et al., 2024), the atmospheric temperature at ~18 km.

For the sub-cloud atmosphere (10–36 km), the PAH composition is consistent with reactions with concentrated $H_2SO_4$. When assuming similar PAH identities across the cloud tops and sub-cloud atmosphere (coronene, pentacene, and ovalene), the spectral fits showed no changes in coronene and pentacene concentrations but a depletion in ovalene below the clouds (**Fig. 5a**). This trend is consistent with the reactivities in concentrated $H_2SO_4$ (*e.g.*, ~85 wt%), where ovalene rapidly decomposes (likely to a quinone (Handa, 1955)) to yield a violet/brown substance (Dias and Koehler, 1995) (*e.g.*, absorbance in the red), while coronene and pentacene oxidize to yield radical cations (Cataldo et al., 2010) with no major changes in structure. Appreciable red absorbances are indeed observed in the Venera 11 spectrum; though, these bands likely overlap with the absorbances of water vapor and $CO_2$ (Maiorov et al., 2005). Larger PAHs (≥6 rings) could potentially form via the thermal decomposition of aerosol-phase IOM after fallout (Okumura and Mimura, 2011; Sephton et al., 2024).

When considering iron, the cloud top composition (ferric chloride or potentially ferric chloride/sulfate) is consistent with recent microphysics (Karyu et al., 2026) and chemical (Egan et al., 2026) models, which support the delivery of ablated iron into the clouds and subsequent alteration to ferric chloride. Best fits were obtained using cross sections for aqueous ferric chloride, which aligns with an assumed solution/liquid state for the mode 1 particles (**Appendix C**).

At the cloud tops, ferric chloride and sulfates could form via reactions with atmosphere-sourced HCl and $H_2SO_4$, respectively (Bertaux et al., 2007; Zhang et al., 2010). Within and below the clouds, differing ferric sulfates would likely form via reactions with the $H_2SO_4$ aerosols (Jiang et al., 2024; Egan et al., 2025b) and thermal decomposition after fallout. Best fits to the spectra for the decomposed cloud aerosols and sub-cloud atmosphere were obtained using the cross sections for solid-phase ferric sulfates, aligning with the assessed physical states (**Appendix C**).

For the aerosols captured by Vega 1 at ~47–52 km, the iron speciation (acid ferric sulfate) aligns with the proposed Vega 1 descent sequence (Mogul et al., 2025). This sequence likely included initial thermal evaporation of water from the captured aerosols in the Vega 1 collection tube just after exiting the cloud deck, temporary increases in $H_2SO_4$ concentration in the captured aerosols (prior to thermal decomposition of $H_2SO_4$), and precipitation of acid ferric sulfate (at ≥18 km). This proposed formation sequence is consistent with the laboratory synthesis of acid ferric sulfate (Jiang et al., 2024).

For the sub-cloud atmosphere at 10–36 km, the iron speciation (ferric hydroxysulfates) aligns with measurements from the Pioneer Venus Large Probe Neutral Mass Spectrometer (LNMS), which identified hydronium jarosite as the major aerosol-derived iron-bearing compound (Mogul et al., 2025). The LNMS, like ISAV 1 and 2, inadvertently captured cloud aerosols in the inlets at ~51 km, where solid-phase hydronium jarosite formed after the thermal loss of water and $H_2SO_4$ from the captured aerosols (Mogul et al., 2025).

When considering the rates of cosmic influx, the total concentrations at the cloud tops for the PAHs (3.4 ± 0.3 pM; 3.8 ± 0.3 ppb) and iron (15 ± 11 pM; 16 ± 12 ppb) would be achieved in geologically short timelines (1.3 ± 0.7 Myr, carbon; 6.1 ± 3.5 kyr, iron). Interpretations as lower limits suggest influx timelines of at least 600 and 2.6 kyr for carbon and iron, respectively. These short timelines support the hypothesis that measurable abundances of UV-absorbers can be acquired through cosmic dust influx and alteration.

The altitude profiles assembled from the spectral fits (**Fig. 5b**) show substantial molar enrichments in the PAHs (≤40-fold) and iron (≥150-fold) from the cloud tops to the cloud aerosols. This enrichment amounts to a minimum aerosol mass loading of ≥8 mg $m^{-3}$ at ~47–51 km. These combined trends align with the measurements from the PV Large Probe Cloud Particle Size Spectrometer (**Fig. 5c**), which showed major enrichments in mass loading of ~10–100-fold or ~1 to 10–85 mg $m^{-3}$ between ~65 and ~47–52 km (Knollenberg and Hunten, 1980). In the sub-cloud atmosphere, we posit that the UV absorbers may be components of the sub-cloud haze layers that potentially form after aerosol fallout from the clouds. In support, we note that the observed haze layers at ~30–46 km (Titov et al., 2018) and ~3.5–6 km (Kulkarni et al., 2025)) occur at the same altitude ranges where Venera 11 (Maiorov et al., 2005) measured the highest visible absorbances (*e.g.*, 10–36 and 2–19 km).

## 5. Conclusion

Thus, we propose a unified origin for Venus' absorbers, where PAHs and iron compounds arise from cometary dust via stepwise alteration pathways from the mesosphere to the surface. This is analogous to the delivery of cosmic dust to the surfaces of Mars (Eigenbrode et al., 2018) and Earth (Plane, 2012). For the PAHs, the large absorption cross sections (≤$10^5$ $M^{-1}$ $cm^{-1}$ at ~300–400 nm (Malloci et al., 2007)) and thermal and radiation stabilities (Stein and Fahr, 1985; Ehrenfreund et al., 2007) support consideration as the primary UV absorbers, while the iron compounds are the dominant absorbers by mass. The development of a low-uncertainty radiative transfer model for the PAHs and iron will assist in these assessments. Surface sourcing of the PAHs and iron-bearing compounds is unlikely (Mogul et al., 2026), as volcanic outputs of organic carbon and iron at Venus are unconstrained, and because volatile iron halides (which could supply gaseous iron to the clouds) are unstable at Venus surface conditions when compared to ferric oxides and pyrite, which are not volatile. Looking ahead, the presence of PAHs and IOM in Venus' atmosphere may be verified by the autofluorescence and ORIGIN instruments on the planned Morning Star missions (Seager et al., 2022).

### Data Availability

Results from the spectral fits will be available on Zenodo upon publication. Absorbance cross sections for the PAHs are included as a Supplementary data file (**Appendix E**) and were obtained from the Theoretical Spectra Database for Polycyclic Aromatic Hydrocarbons (https://astrochemistry.oa-cagliari.inaf.it/database/pahs.html). Spectra and cross sections for the iron compounds were obtained from the referenced reports. Venus spectra were obtained from the referenced reports.

### Code Availability

No code was generated for this work. All data fitting procedures were conducted in Microsoft Excel.

### Author Contributions

**Rakesh Mogul:** Conceptualization, Formal Analysis, Investigation, Methodology, Visualization, Writing – original draft, Writing – Review & Editing. **Mikhail Zolotov:** Conceptualization, Writing – Review & Editing. **Michael Way:** Writing – Review & Editing. **Sanjay Limaye:** Writing – Review & Editing.

### Ethics Declarations

Authors declare they have no competing interests.

**Acknowledgements**
RM, MYZ, and SSL acknowledge administrative support from Blue Marble Space. RM thanks Santiago Pérez Hoyos and Kandis-Lea Jessup for helpful discussions.

**Funding**
RM, MYZ, MJW, and SSL acknowledge support from the NASA Solar Systems Working program (award 80NSSC24K0929). MYZ was also supported by the NASA Discovery program (award 80NSSC22M0187). MJW was also supported through the NASA Astrobiology Program, NExSS, NASA Habitable Worlds Program, and the GSFC Sellers Exoplanet Environments Collaboration.

## Appendix A. *Support for a multi-component absorption model*

A multi-component absorption model using the candidates in **Table 1** was supported by the trends in the $\chi^2_{red}$ and AIC terms. The impacts of multiple PAHs and the inclusion of iron-bearing compounds on the $\chi^2_{red}$ and AIC terms are shown in **Appendix Figs. D.6** and **D.7**. For the cloud tops, (1) increasing the number of PAHs in the model from 1 to 3 decreased the $\chi^2_{red}$ from 150 to ~1 (**Appendix Fig. D.7A**) and (2) the AIC substantially minimized (ΔAIC ≈ 180) when the number of PAHs increased from 2 to 3 (**Appendix Fig. D.6A**). This suggested that 3 PAHs better fit the cloud top spectrum and uncertainties and that 1–2 PAHs yielded poor fits (among the candidates in **Table 1**). For the cloud aerosols, (1) increasing the number of PAHs from 1 to 4 decreased the $\chi^2_{red}$ from 26 to ~2 (**Appendix Fig. D.7B**) and (2) the AIC substantially minimized (ΔAIC ≈ 90) when the number of PAHs increased from 3 to 4 (**Appendix Fig. D.6C**). This suggested that better fits to the aerosols were obtained using multiple absorbers (among the candidates in **Table 1**). The fit qualities for the cloud tops (upper limit) and cloud aerosols were not negatively affected by increasing the number of parameters. Rather, the lowest AIC values were obtained with the highest number of fitted parameters (**Appendix Figs. D.6A** and **D.6C**). In contrast, the fit qualities for the cloud tops (lower limit) (ΔAIC ≈ 8) and sub-cloud atmosphere (ΔAIC ≈ 3) were indeed negatively impacted, though marginally, with increased AIC values (**Appendix Figs. D.6B** and **D.6D**). Given the low ΔAIC, the spectral fit with 2 PAHs + FC was treated as a plausible alternative to 1 PAH.

## Appendix B. *Interpretation of the spectral fits and associated uncertainties*

Several knowledge gaps regarding the physical states, particle sizes, and spectra of the PAHs and iron-bearing compounds could not be accounted for in this study. These knowledge gaps and the associated uncertainties, as described further below, present several challenges towards developing a low-uncertainty radiative transfer model for a mixture of PAHs and iron in the cloud tops, decomposed cloud aerosols, and sub-cloud atmosphere. Therefore, we conducted a first order comparison of PAHs and iron-bearing compounds to the Venus spectra and accounted for the knowledge gaps, in the absence of a robust radiative transfer model, by treating the spectral fitting results, such as the PAH identities and exact concentrations, as proxies for the *total* PAH concentration, *total* concentration of the PAH-associated carbon, and the general properties of the PAHs, including the number of rings, reactivity trends in $H_2SO_4$, thermal stabilities, and red shifts in the absorbance. In turn, the *total* PAH and iron concentrations were treated as lower limits. This assessment is based on the slightly lower iron concentration obtained from our results (80 ± 10 pM), when fitting the cloud top spectra with only iron (**Appendix Figs. D.1d** and **D.2d**), compared to the estimated iron concentration (~100 pM) in Venus' clouds from the radiative transfer model for ferric chloride in Egan et al. (2026). The associated timelines of influx for carbon and iron were also treated as lower limits. Summarized in the following bullet list are relevant knowledge gaps and associated uncertainties for the spectral model.

- Databases of absorption cross sections were not readily available for PAHs that (a) were obtained at varying particle/grain sizes, film thicknesses, and temperatures, (b) included PAHs bearing alkyl, carbonyl, and/or heteroatom groups, and (c) sufficiently overlapped with the Venus spectra in the UV *and* visible wavelengths. In these absences, the Venus spectra (240–650 nm) were compared to an internally consistent collection of cross sections for PAHs at the ground state (*e.g.*, single molecule in the gas phase) from the TSD-PAHs (Malloci et al., 2007). These cross sections are similar to those used in the literature to model interstellar and exo-atmospheric PAHs (*e.g.*, Draine and Li (2007) and Arenales-Lope et al. (2025)).

- The respective absorption spectra for tetracene, coronene, and ovalene are noticeably different (*e.g.*, changes in peak center and width) between the ground state and thin films with thicknesses of 0.10–0.45 µm (Bryson et al., 2011; Maddii Fabiani et al., 2024). These differences partly arise from non-covalent bonding between the stacked PAHs in the films (Maddii Fabiani et al., 2024). This is relevant since particles of PAHs with diameters of ≥0.1 µm may exhibit similar absorption features. However, under Venus conditions, the sizes and abundances of the particles containing PAHs and/or iron compounds remain unconstrained. This includes particles that form above and at the cloud tops, particles that released from the ISAV 1 mirrors, and particles that formed after aerosol fallout below the clouds. These uncertainties, along with the lack of available absorption cross sections for PAHs at differing particle sizes, precluded a systematic assessment of particle size in this study.

- For the iron-bearing compounds, the cross sections were obtained from Jiang et al. (2024), Egan et al. (2025b), and Egan et al. (2026), which modeled the absorption of ferric compounds against Venus' cloud top spectra. Unlike the PAHs, these cross sections did not represent the ground-state; rather, the cross sections were experimentally determined at room temperature using different procedures (El-Hosainy et al., 2021; Jiang et al., 2024; Egan et al., 2025b), including the assumptions described in the **Section 2.1.3** for the ferric hydroxysulfates. For ferric chloride and ferric sulfates, absorbance cross sections at differing particle sizes and temperatures were not readily available.

- The compositional heterogeneity (*e.g.*, relative mass fractions) of particles that potentially contain PAHs, iron, and/or sulfuric acid is unconstrained. The results from this study suggest that PAHs are minor components by mass across the atmosphere. This is relevant since the spectroscopic properties of particles containing iron, sulfuric acid, and minor abundances of PAHs are not well known; where the uncertainties include the degree of stacking by the PAHs, particle size, and compositional heterogeneity. Similarly, radiative transfer models for iron in Venus' atmosphere remain understudied (*e.g.*, Egan et al. (2026)). In this context, we conducted a first order comparison of PAHs and iron-bearing compounds to the Venus spectra, consistent with the analytical procedures in Pérez-Hoyos et al. (2018), Jiang et al. (2024), Egan et al. (2025b), and others.

- The sizes of particles containing organic matter at the cloud tops are unconstrained. The potential upper and lower limits could be ~100 and <1 μm, respectively. For organic matter that survives ablation at Venus, particles with diameters of ≤100 μm exhibit the highest assessed influx rates (Carrillo-Sánchez et al., 2020). However, in the upper clouds, the particles with the highest concentrations have diameters of ≤1 μm, as measured by the Pioneer Venus Large Probe Cloud Particle Size Spectrometer (LCPS) (Knollenberg and Hunten, 1980). This wide range of potential particle sizes for the PAHs introduces considerable uncertainty when accounting for scattering and transmittance.

- When considering temperature, the MPI-Mainz UV/Vis Spectral Atlas (Keller-Rudek et al., 2013) contains spectra for ~6 PAHs at temperatures relevant to Venus' sub-cloud atmosphere (*e.g.*, 10–36 km, ~450–660 K). These spectra were obtained in the ultraviolet wavelengths at ~200–400 nm, but not in the visible wavelengths. For anthracene vapor, compared to the ground state, these spectra (~300–400 nm) showed maximum increases (up to ~4-fold) in absorbances at 433 K, followed by peak broadening at ≥573 K. Unfortunately, the pertinent spectra (<400 nm) in the MPI-Mainz UV/Vis Spectral Atlas do not overlap with the Venera 11 spectra for the sub-cloud atmosphere, which were obtained in the visible wavelengths (>400 nm). Therefore, this precluded a systematic assessment of temperature using the MPI-Mainz UV/Vis Spectral Atlas.

**Appendix C.** ***Assessing the physical states of the PAHs and iron-bearing compounds***

The physical states of the PAHs were considered as follows. At the atmospheric conditions at 75 ± 7 km (190–235 K, 0.003–0.05 bar (Seiff et al., 1985)), or at the cloud tops, the identified PAHs (from the fitting procedures) were assumed to be gases or potentially components of the mode 1 particles, where the compositional heterogeneity of the mode 1 particles was unconstrained. At the atmospheric conditions at 18 km (~600 K, 26 bar (Seiff et al., 1985)), or within the Vega 1 collection tube (≤600 K), the identified PAHs were assumed to be single molecules or small nanoparticles (<100 nm) that were released from the mirrors after the mechanical shock. In sub-cloud atmosphere, at the conditions at 10–36 km (~450–660 K, ~5–47 bar (Seiff et al., 1985)), the identified PAHs were assumed to be gases or nanoparticles. As described, our spectral fits relied on the ground-state cross sections because suitable spectral databases for PAHs at differing temperatures and particle sizes were not available.

For the iron compounds, the physical state at the cloud tops was presumed to be the solution phase, presumably in the mode 1 particles. Among the solution and solid-phase candidates, the best fits to the cloud tops were obtained when using the cross sections for aqueous solutions. The iron compounds from the partly decomposed aerosols and sub-cloud atmosphere were assumed to be in the solid phase. Among the solution and solid-phase candidates, the best fits to the partly decomposed aerosols and sub-cloud atmosphere were obtained when using the cross sections for the solid phase samples.

**Appendix D.** ***Supporting Figures and Table***

***Figure. D.1***

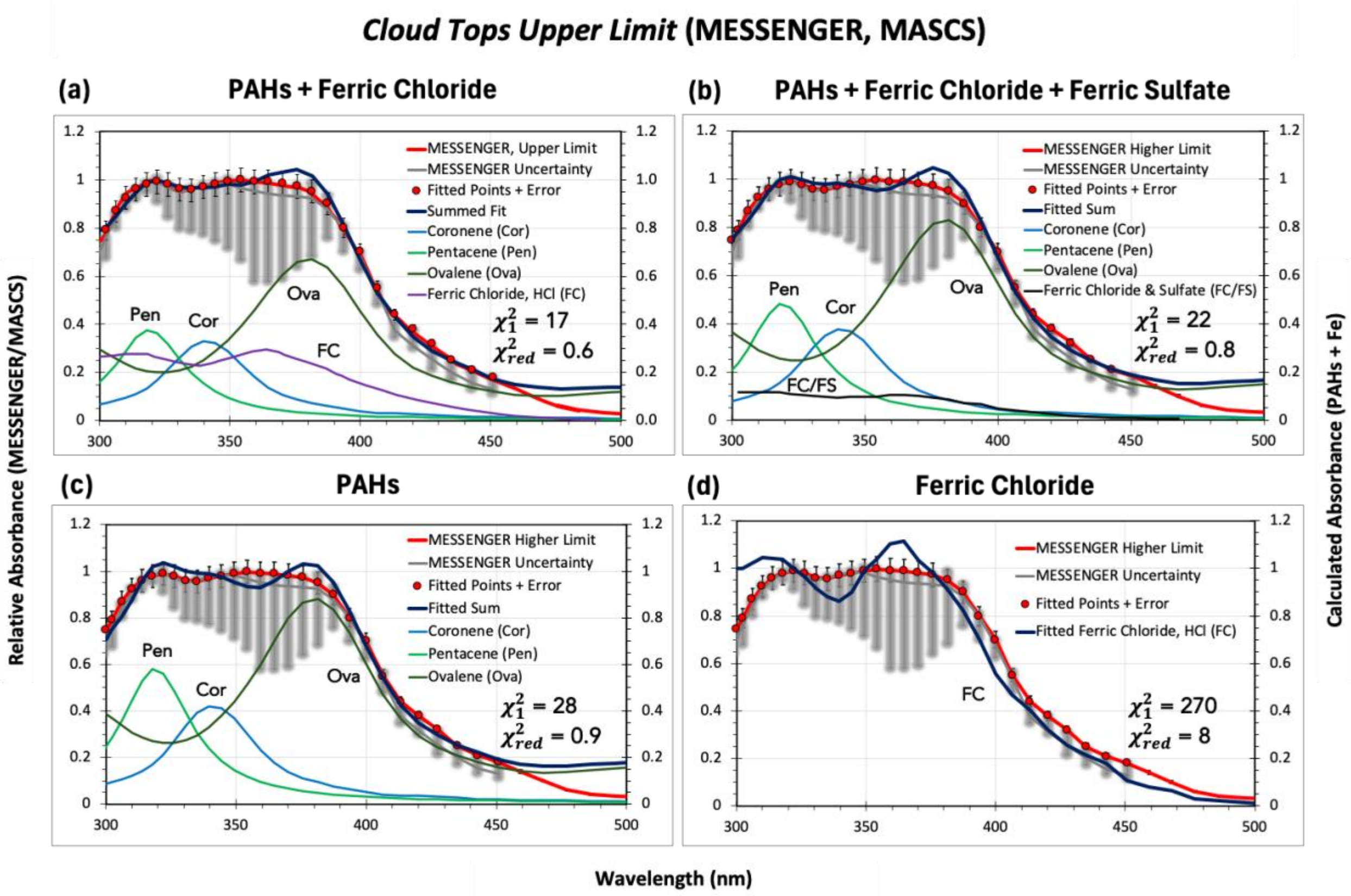

**Figure D.1.** Spectral fits to the maximum relative absorbance of the cloud tops. The absorbance spectrum (75 ± 7 km) was measured by MESSENGER/MASCS (left-hand y-axis) and fitted with differing mixtures of PAHs and iron compounds (right-hand y-axis), which are described in the plot titles and legends. The fits are ordered using the chi-squared statistic ($\chi_1^2$) and organized from best (panel a) to worst (panel d). The red lines are the maximum cloud top absorbance, red circles are the fitted data points, gray lines are the reported optimal absorbance, and the blue lines are the summed absorbance of the mixtures. Gray error bars are the reported maximum and minimum uncertainty limits, and black error bars represent the reported relative error in the measurement (5%).

**Appendix D.**

***Figure. D.2***

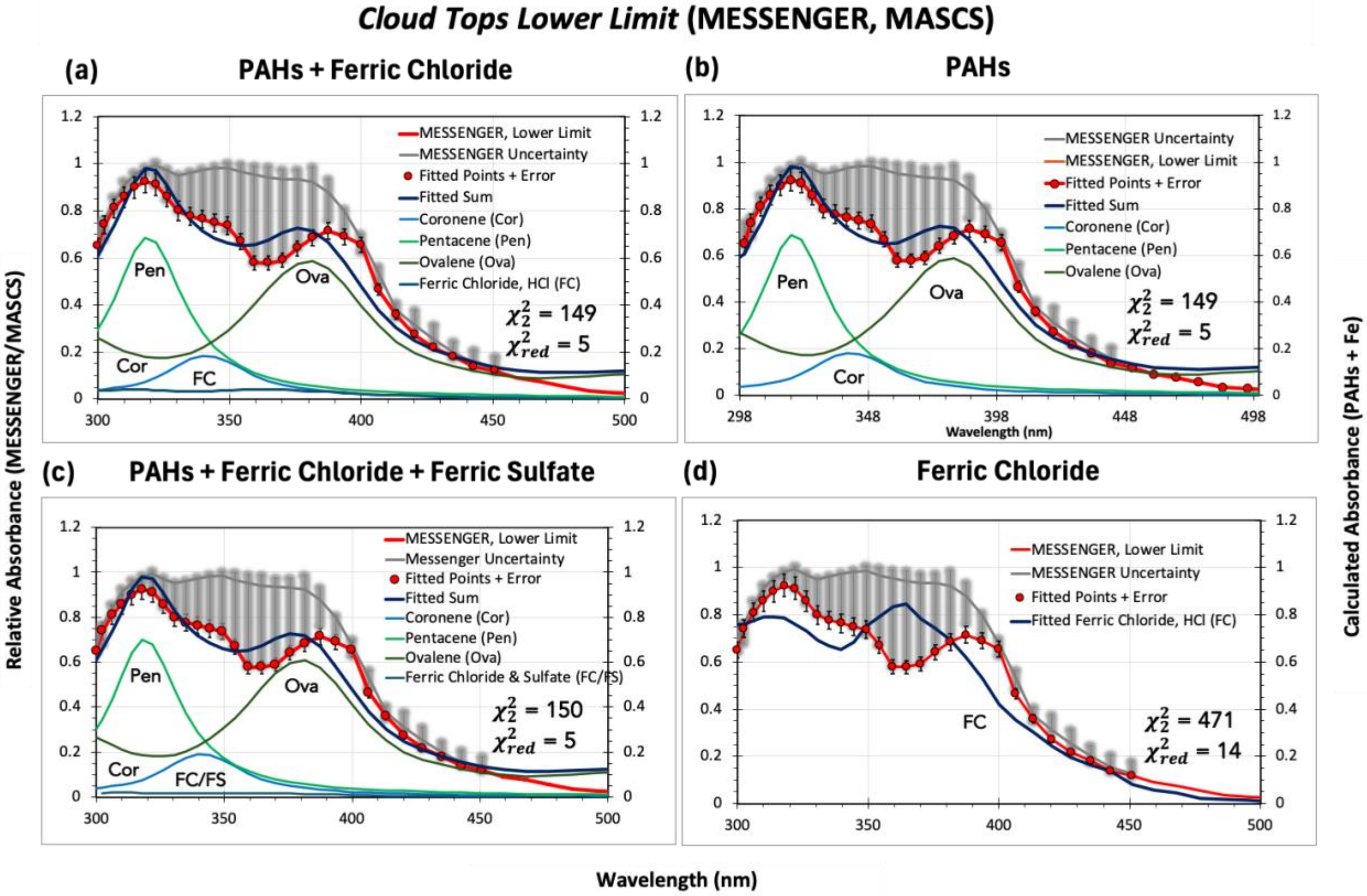

**Figure D.2.** Spectral fits to the minimum relative absorbance of the cloud tops. The absorbance spectrum (75 ± 7 km) was measured by MESSENGER/MASCS (left-hand y-axis) and fitted with differing mixtures of PAHs and iron compounds (right-hand y-axis), which are described in the plot titles and legends. The fits are ordered using the chi-squared statistic ($\chi^2_2$) and organized from best (panel a) to worst (panel d). The red lines are the minimum cloud top absorbance, red circles are the fitted data points, gray lines are the reported optimal absorbance, and the blue lines are the summed absorbance of the mixtures. Gray error bars are the reported maximum and minimum uncertainty limits, and black error bars represent the reported relative error in the measurement (5%).

**Appendix D.** ***Figure. D.3***

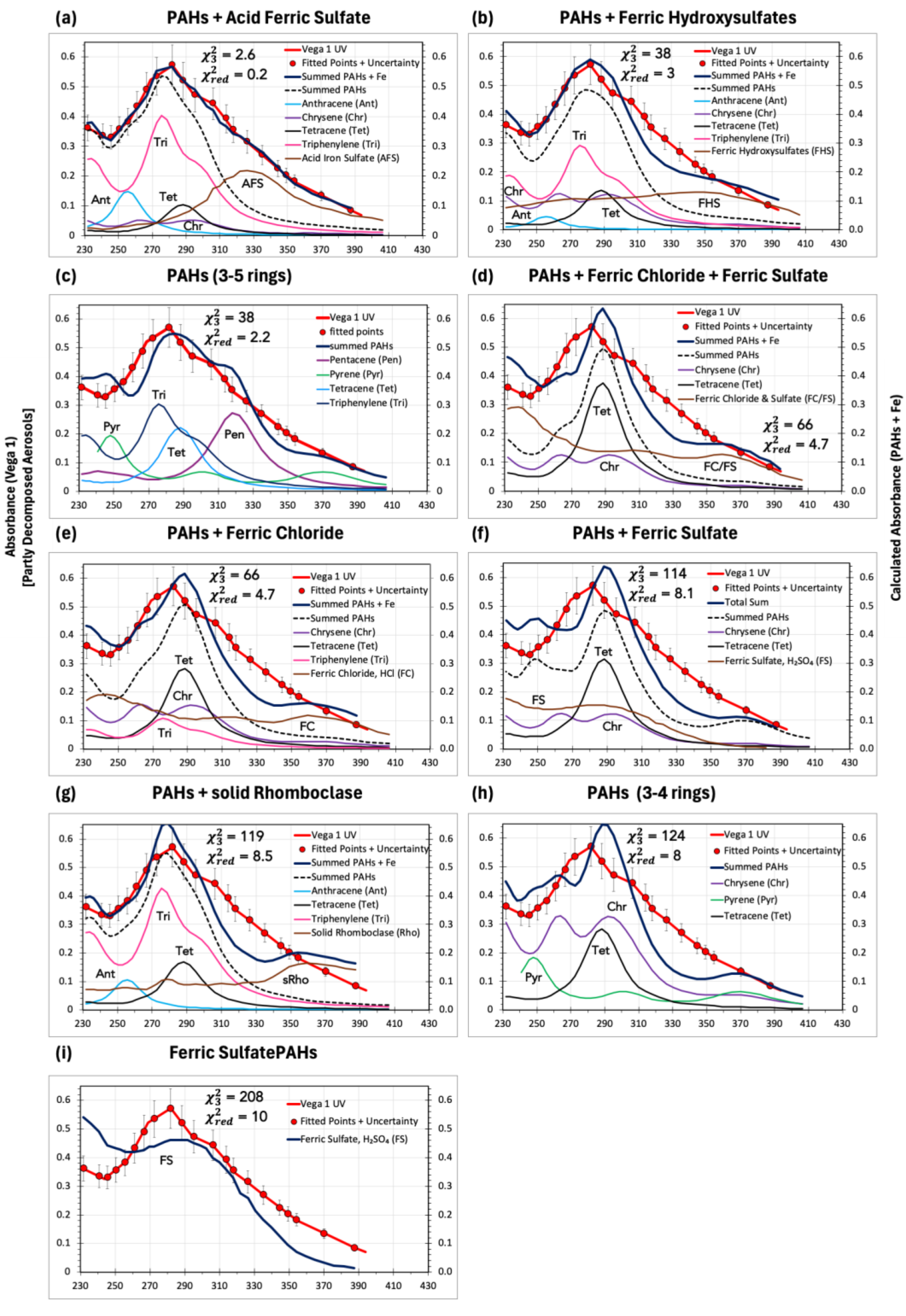

**Figure D.3.** Spectral fits to the absorbance of the partly decomposed aerosols. The absorbance spectrum was measured at 18 km by the ISAV 1 instrument on Vega 1 (left-hand y-axis) and fitted with differing mixtures of PAHs and iron compounds (right-hand y-axis), which are described in the plot titles and legends. The fits are ordered using the chi-squared statistic ($\chi^2_3$) and organized from best (panel a) to worst (panel i). The red lines are the aerosol absorbance, red circles are the fitted data points, the blue lines are the summed absorbance of the mixtures, and the dashed lines are the summed absorbance for just the PAHs. Black error bars represent the relative uncertainty in the measurement (12%). Fits using 3–5 ring PAHs are provided for comparison purposes in panel (c); the 5 ring PAHs were not considered as viable candidates given their lower vapor pressures and presumed lower yields of release from the ISAV mirrors during the mechanical shock at ~18 km.

**Appendix D.**

***Figure. D.4***

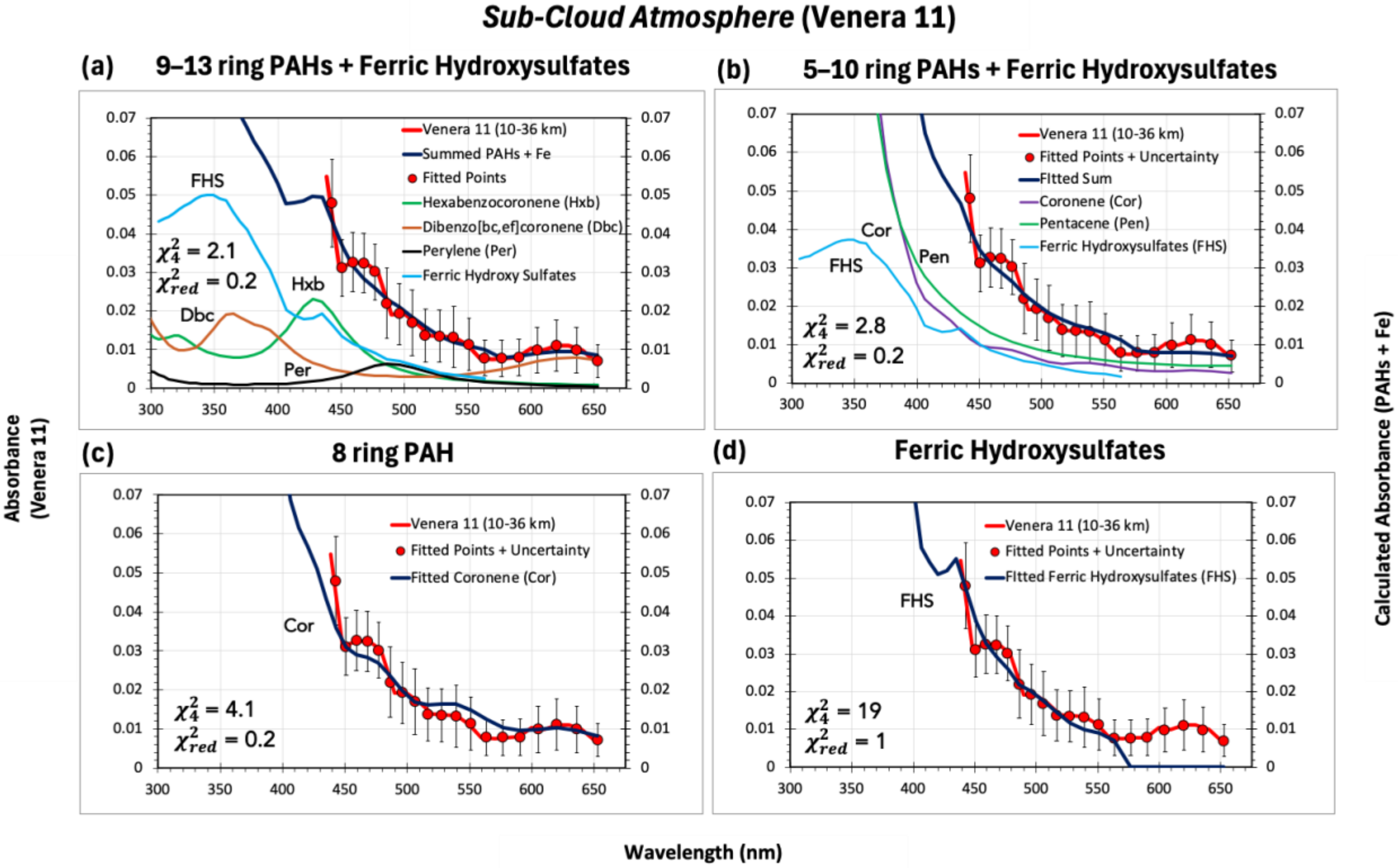

**Figure D.4.** Spectral fits to the sub-cloud atmosphere absorbance. The absorbance spectrum was measured at 10-36 km by the spectrophotometer on Venera 11 (left-hand y-axis) and fitted using differing mixtures of PAHs and iron compounds (right-hand y-axis), which are described in the plot titles and legends. The fits are ordered using the chi-squared statistic ($\chi^2_4$) and organized from best (panel A) to worst (panel D). The red lines are the sub-cloud atmosphere absorbance, red circles are the fitted data points, and the blue lines are the summed absorbance of the mixtures. Black error bars represent the relative uncertainty in the measurement (~41%).

## Appendix D.

### *Figure. D.5*

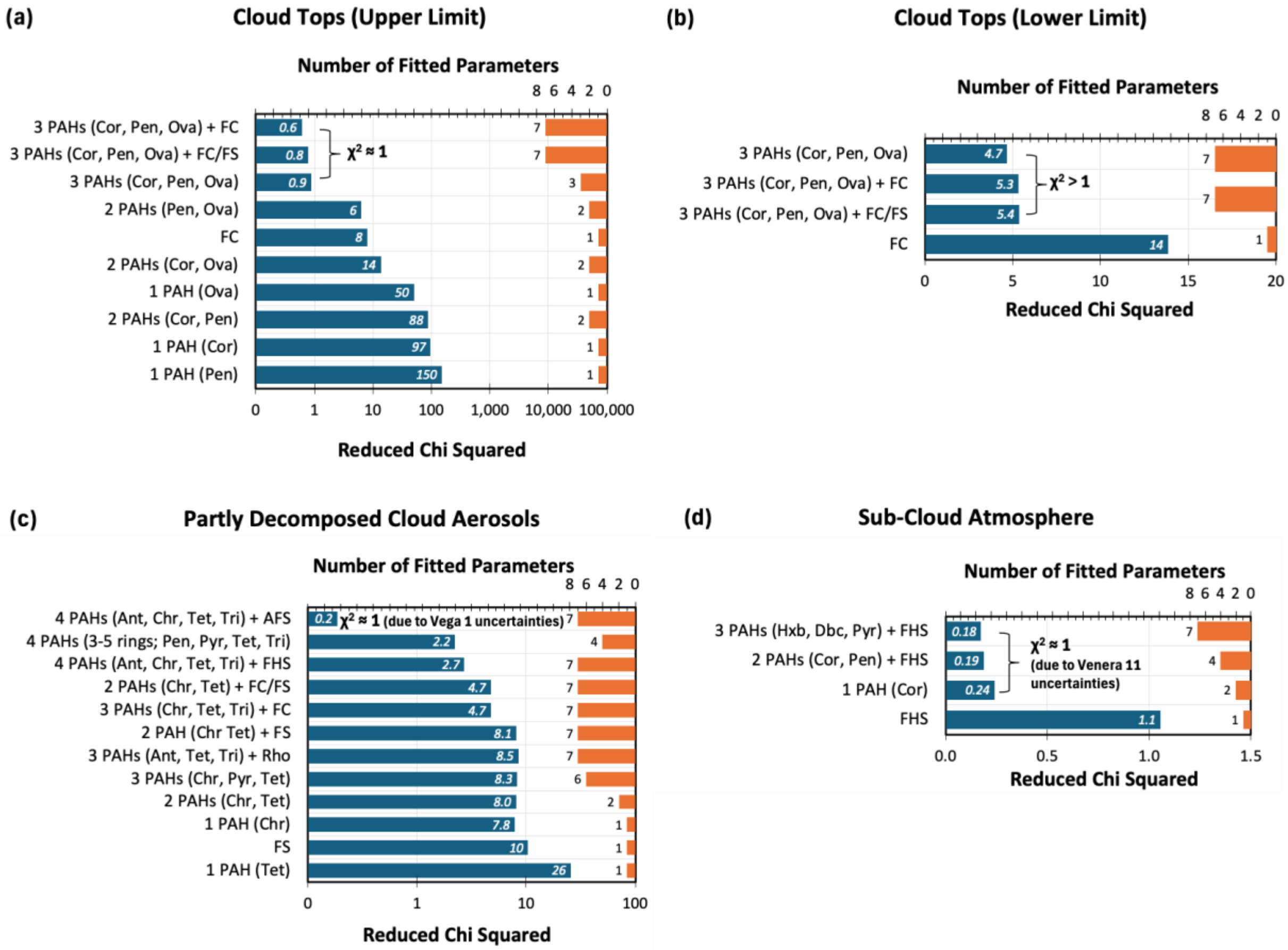

**Figure D.5.** Trends in the reduced chi squared statistic ($\chi^2_{red}$) when using different combinations of PAHs and iron compounds. Relationships between composition (left y-axis) and number of initial fitting parameters (right y-axis; upper x-axis) with the $\chi^2_{red}$ are shown for the (a) cloud tops (upper limit), (b) cloud top (lower limit), (c) partly decomposed aerosols, and (d) sub-cloud atmosphere. Spectral fits with $\chi^2_{red} \approx 1$ are noted and the adjustments rationalized to account for the high measurement uncertainties for the Vega 1 and Venera 11 spectra.

## Appendix D.

*Figure. D.6*

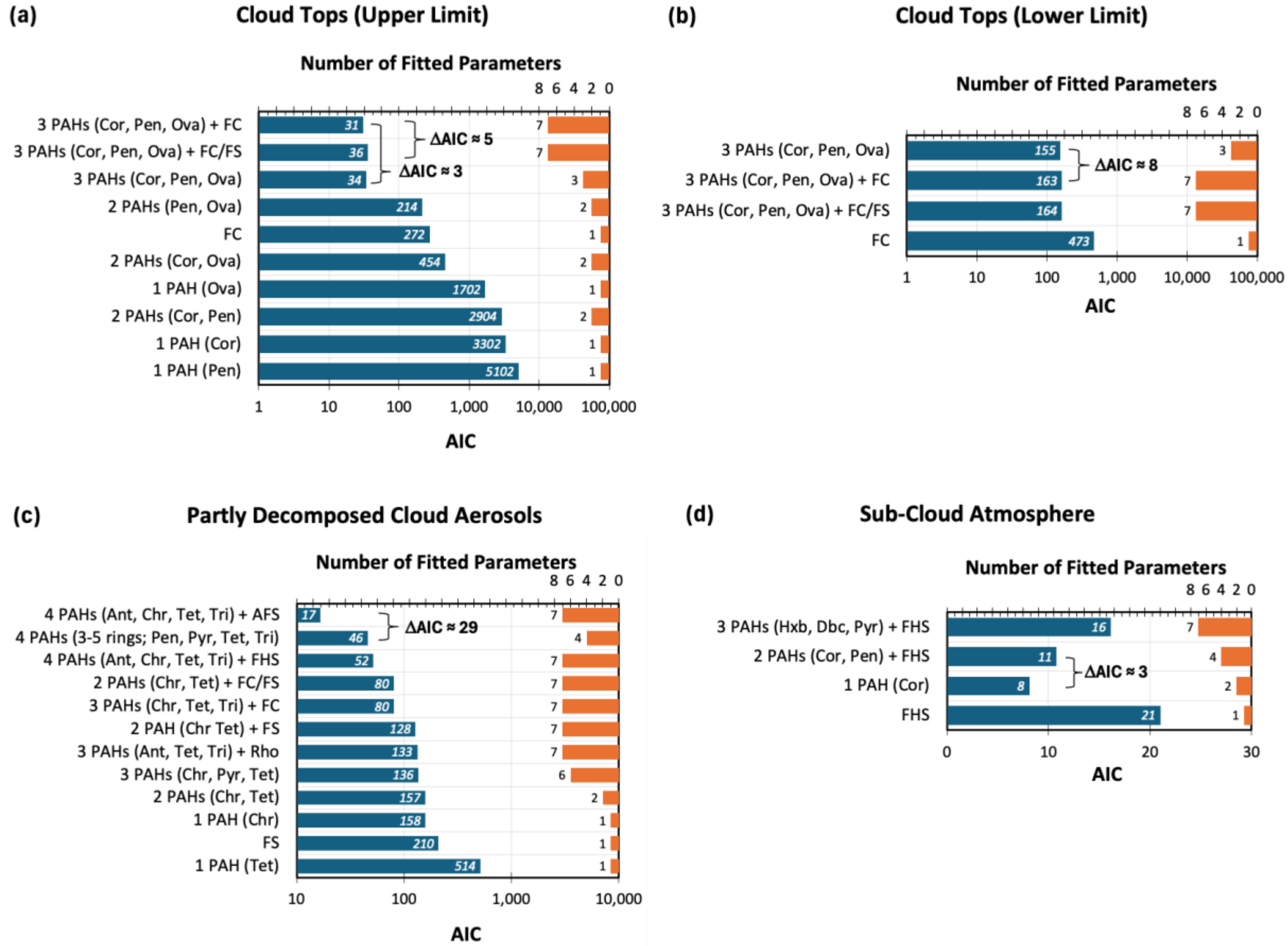

**Figure D.6.** Trends in the Akaiki Information Criterion (AIC) when using different combinations of PAHs and iron compounds. Relationships between composition (left y-axis) and number of initial fitting parameters (right y-axis; upper x-axis) with the AIC are shown for the (a) cloud tops (upper limit), (b) cloud top (lower limit), (c) partly decomposed aerosols, and (d) sub-cloud atmosphere. The ΔAIC are calculated against the lowest respective AIC value.

**Appendix D.**

*Figure. D.7*

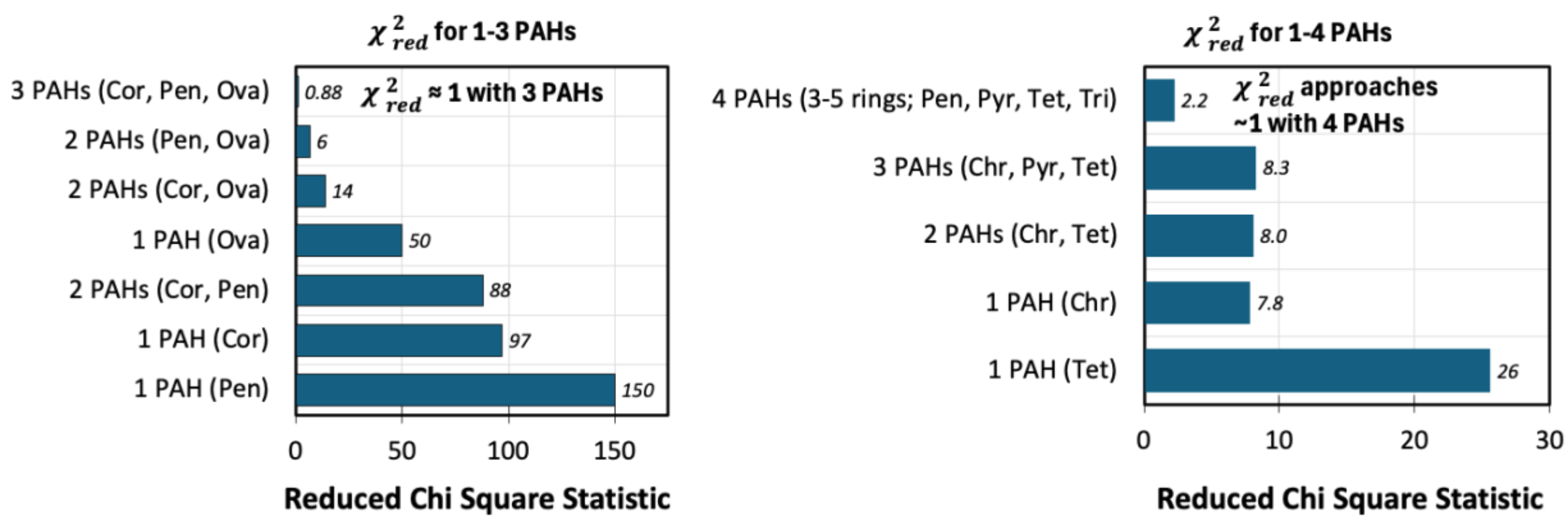

**Figure D.7.** Trends in the reduced chi squared statistic ($\chi^2_{red}$) when fitting the (a) cloud tops with 1-3 PAHs and (b) cloud aerosols with 1-4 PAHs. For the cloud tops, $\chi^2_{red} \approx 1$ when using 3 PAHs. For the cloud aerosols, $\chi^2_{red}$ approaches ~1 when using 4 PAHs.

**Appendix D.**

***Table. D.1***

**Table D.1.** Molecular compositions that best fit the Venus UV-blue spectra. Composition for the captured aerosols are provided before and after normalization using the LNMS aerosol iron concentration.

| Spectra | | PAHs | | | | | Iron |
|---|---|---|---|---|---|---|---|
| Cloud Tops | Altitudes (km) | Total PAHs (pM) | Pentacene (pM) | Coronene (pM) | | Ovalene (pM) | Ferric Chloride (pM) |
| | 75 ± 7 | 3.4 ± 0.3 (5–10 rings) | 0.5 ± 0.2 | 0.8 ± 0.2 | | 2.1 ± 1.4 | 14 ± 11 |
| Sub-Cloud Atmosphere | Altitudes (km) | Total PAHs (pM) | Pentacene (pM) | Coronene (pM) | | Ovalene (pM) | Ferric Hydoxysulfates (pM) |
| | ~52–47 | 1.7 ± 0.7 (5–10 rings) | 0.8 ± 0.3 | 0.9 ± 0.4 | | 0 | 83 ± 34 |
| Partly Decomposed Aerosols | Altitudes (km) | Total PAHs (nM) | Anthracene (nM) | Chyrsene (nM) | Tetracene (nM) | Triphenlene (nM) | Acid Ferric Sulfate (nM) |
| | 36–10 | 39 ± 5 (3–4 rings) | 4.8 ± 0.6 | 4.0 ± 0.5 | 2.6 ± 0.3 | 27 ± 3 | 6500 ± 800 |
| Partly Decomposed Aerosols (*normalized*) | Altitudes (km) | Total PAHs (pM) | Anthracene (pM) | Chyrsene (pM) | Tetracene (pM) | Triphenlene (pM) | Acid Ferric Sulfate (pM) |
| | 36–10 | 94 ± 52 (3–4 rings) | 39 ± 5 | 4.8 ± 0.6 | 4.0 ± 0.5 | 2.6 ± 0.3 | 16 ± 12 |

## Appendix E. *Supplementary Data*

- Absorbance cross sections for the PAHs are included as a Supplementary data file (.xslx) and were obtained from the Theoretical Spectra Database for Polycyclic Aromatic Hydrocarbons (https://astrochemistry.oa-cagliari.inaf.it/database/pahs.html).